\documentclass[aps,pre,superscriptaddress,nofootinbib,twocolumn,a4paper]{revtex4-2}

\usepackage{newlfont}
\usepackage[english]{babel}
\usepackage{amsfonts}
\usepackage{graphicx}
\usepackage[toc]{appendix}
\usepackage[caption=false]{subfig}
\usepackage{tensor}
\usepackage{nicefrac}
\usepackage{booktabs}
\usepackage{calc}
\usepackage{siunitx}
\usepackage{placeins}
\usepackage{textcomp}
\usepackage{listings}
\usepackage[T1]{fontenc}
\usepackage[utf8]{inputenc}
\usepackage{amssymb}
\usepackage{tabularx}
\usepackage{array}
\usepackage{hhline}
\usepackage{amsthm}
\usepackage[fleqn]{amsmath}
\usepackage{mathtools}
\usepackage{amssymb}
\usepackage{calligra}
\usepackage{bbold}
\usepackage{dsfont}
\usepackage{braket}
\usepackage{hhline}
\usepackage{easyReview}
\usepackage{enumitem}
\usepackage{natbib}
\usepackage{csquotes}
\usepackage{color}
\usepackage[bookmarks=true, colorlinks=true, urlcolor=blue, citecolor=blue, linkcolor=blue, pdfborder={0 0 0}]{hyperref}
\usepackage{bookmark}
\usepackage{todonotes}
\usepackage{easyReview}

\newlength{\mintednumbersep}
\AtBeginDocument{%
  \sbox0{\tiny00}%
  \setlength\mintednumbersep{\parindent}%
  \addtolength\mintednumbersep{-\wd0}%
}

\definecolor{cadmiumgreen}{rgb}{0.0, 0.42, 0.24}
\definecolor{sigmoid_red}{HTML}{E93628}
\definecolor{sigmoid_blue}{HTML}{00A2FF}
\definecolor{sigmoid_black}{HTML}{000000}
\definecolor{TMNTpurple}{HTML}{AA1BDD}
\definecolor{TMNTblue}{HTML}{00AAE6}
\definecolor{TMNTyellow}{HTML}{FFA500}
\definecolor{TMNTred}{HTML}{E3331C}

\newcommand{\edge}{\varepsilon}
\newcommand{\walkerstep}{\eta}

\newcommand{\group}{\mathcal{G}}
\newcommand{\loops}{t}
\newcommand{\prob}{p}

\newcommand{\probmetareinf}{\prob_{mr}}

\newcommand\bsfrac[2]{%
\scalebox{-1}[1]{\nicefrac{\scalebox{-1}[1]{$#1$}}{\scalebox{-1}[1]{$#2$}}}%
}
\newcommand{\savg}[1]{\langle #1 \rangle_s}

\begin{document}

\title{Meta-plasticity and memory in multi-level recurrent feed-forward networks}
%
\author{Gianmarco Zanardi}
\affiliation{Physics Department, University of Trento, via Sommarive, 14 I-38123 Trento (IT)}
\affiliation{INFN-TIFPA, Trento Institute for Fundamental Physics and Applications, I-38123 Trento (IT)}
\author{Paolo Bettotti}
\affiliation{Physics Department, University of Trento, via Sommarive, 14 I-38123 Trento (IT)}
\author{Jules Morand}

\affiliation{Department of Civil, Environmental and Mechanical Engineering, University of Trento, via Mesiano 77, I-38123, Trento, Italy}
\affiliation{INFN-TIFPA, Trento Institute for Fundamental Physics and Applications, I-38123 Trento (IT)}
\author{Lorenzo Pavesi}
\affiliation{Physics Department, University of Trento, via Sommarive, 14 I-38123 Trento (IT)}
\author{Luca Tubiana}
\email[Corresponding author:]{luca.tubiana@unitn.it}
\affiliation{Physics Department, University of Trento, via Sommarive, 14 I-38123 Trento (IT)}
\affiliation{INFN-TIFPA, Trento Institute for Fundamental Physics and Applications, I-38123 Trento (IT)}
\begin{abstract}
Network systems can exhibit memory effects in which the interactions  between different pairs of nodes adapt in time, leading to the emergence of preferred connections, patterns, and sub-networks.
To a first approximation, this memory can be modelled through a ``plastic'' Hebbian or homophily mechanism, in which edges get reinforced proportionally to the amount of information flowing through them.  However, recent studies on glia-neuron networks have highlighted how memory can evolve due to more complex dynamics, including multi-level network structures and ``meta-plastic'' effects that modulate reinforcement.
Inspired by those systems, here we develop a simple and general model for the dynamics of an adaptive network with an additional meta-plastic mechanism that varies the rate of Hebbian strengthening of its edge connections. The meta-plastic term acts on a second network level in which edges are grouped together, simulating local, longer time-scale effects.
Specifically, we consider a biased random walk on a cyclic feed-forward network. The random walk chooses its steps according to the weights of the network edges. The weights evolve through a Hebbian mechanism modulated by a meta-plastic reinforcement, biasing the walker to prefer edges that have been already explored. 
We  study the dynamical emergence (memorisation) of preferred paths and their retrieval and identify three regimes: one dominated by the Hebbian term, one in which the meta-reinforcement drives memory formation, and a balanced one. We show that, in the latter two regimes, meta-reinforcement allows the retrieval of a previously stored path even after the weights have been reset to zero to erase Hebbian memory. 
\end{abstract}
%
\maketitle
\section{Introduction}
\label{sec:introduction}
Several biological, ecological and social systems, such as for example internet communities, human populations, epigenetic interactions, and ecosystems, can be described as networks of interacting agents~\cite{strogatz_exploring_2001, wuchty_architecture_2006, castellano_statistical_2009, sneppen_simplified_2010, mendez_stochastic_2014, pastor-satorras_epidemic_2015, treur_network-oriented_2016, dawid_agent-based_2018}.
These networks are in general adaptive: their topology and the strength of their connections evolve in time in response to the interaction between the agents~\cite{treur_network-oriented_2020, sawicki_perspectives_2023}. 
This gives rise to the emergence of transitions and collective behaviours such as network polarisation and opinion dynamics \cite{kozma_consensus_2008, sirbu_opinion_2017, baumann_modeling_2020}, epidemic spreading \cite{gross_epidemic_2006, giordano_modelling_2020, morand_quality_2024} or mobility and diffusion patterns \cite{gonzalez_understanding_2008, rosvall_memory_2014}.

Network dynamics often display memory effects, where a preference for certain connections or paths emerges over time as a result of interactions amongst the agents or external effects: this is the case for example of communities and echo chambers in social networks~\cite{gaumont_reconstruction_2018, lorenz-spreen_tracking_2018, baumann_modeling_2020, cinelli_echo_2021}, preferred patterns mobility networks~\cite{rosvall_memory_2014, vilk_phase_2022} and, of course, memories in the brain~\cite{hopfield_neural_1982, hopfield_neurons_1984, van_essen_wu-minn_2013}.

The brain provides a remarkable example of adaptive and memorising network, with multiple forms of memory across different time-scales~\cite{james_memory_1890, cowan_chapter_2008}.
Neurons can be considered as agents connected by synapses that exchange neuro-transmitter molecules to communicate.
The variation in strength of specific synapses, called \textit{plasticity}, leads to the formation of patterns of neurons that activate together, which are assumed to encode our memories~\cite{semon_mneme_1911, hebb_organization_1949, josselyn_finding_2015, josselyn_memory_2020}, as modelled by Hopfield~\cite{hopfield_neural_1982,hopfield_neurons_1984}. 

The first notable theoretical framework of plasticity was presented by~\textcite{hebb_organization_1949}:
it can be summarised with the famous 
sentence \guillemotleft neurons that fire together wire together\guillemotright. Hebbian theory states that when two neurons repeatedly activate together their synapses become stronger, making it easier to happen again in the future.
This very fundamental description of plastic synapses provides the basis for numerous more advanced memory models~ \cite{bienenstock_theory_1982, easton_stabilization_1984, gerstner_neuronal_1996, clopath_connectivity_2010}. 
In this context, the susceptibility of a synapse to the activity of the neurons it connects is called \textit{learning rate}:
the higher the rate, the greater the strengthening of the synapse when a signal traverses it.

Analogous concepts to plasticity exist for adaptive networks other than the brain, e.g. homophily bonding~\cite{mcpherson_birds_2001, treur_network-oriented_2020}, mutualistic/antagonistic interactions~\cite{hui_modelling_2018} and prey/predator dynamics~\cite{knebel_coexistence_2013} in social, economic or ecological networks respectively.
These concepts provide the bases for emergent network properties and introduce non-Markovianity in the dynamics, effectively introducing a memory kernel for the process~\cite{baronchelli_nonequilibrium_2007, rosvall_memory_2014, matamalas_assessing_2016}.

In the last two decades, it has also become increasingly clear that plasticity in the brain is further modulated by glial cells, including astrocytes, that interact chemically with synapses~\cite{araque_tripartite_1999, newman_new_2003, de_pitta_computational_2012, de_pitta_astrocytes_2016, pajevic_oligodendrocyte-mediated_2023} over time-scales that can be larger compared to the intrinsic ones of neurons~\cite{araque_gliotransmitters_2014}.
The interaction between glial cells and neurons has been described either in terms of tripartite synapses, or of  ``multiplex'' networks~\cite{de_domenico_mathematical_2013}, i.e. networks composed of two levels: the neuronal one and the glial one, which constitute two sub-networks interacting with each other.
Multi-level structures such as the glial-neuron system can also be 
found in other systems, such as interconnected social networks, ecological networks or bio-physical and bio-molecular networks~\cite{de_domenico_mathematical_2013, dickison_multilayer_2016, pilosof_multilayer_2017, de_domenico_more_2023, mukeriia_multi-adaptive_2024}.

One possible effect of the interaction between glial cells and neurons is to vary the rate of synaptic reinforcement modifying their learning rate.
This \textit{meta-plastic} reinforcement has been studied in theoretical models~\cite{fusi_cascade_2005,yger_models_2015}.
In particular, \textcite{fusi_cascade_2005} employed a mean-field approach to study how the inclusion of meta-plasticity in a model of binary synapses helps preserve memory in terms of signal-to-noise ratio.
Arguably, similar dynamics should act in other kind of network systems, for example in social networks by priming the susceptibility of agents to information coming from trusted connections.
Recently, learnable activation functions are also being explored in novel Neural Network architectures named Kolmogorov-Arnold networks~\cite{liu_kan_2024}.

The interplay between plastic and meta-plastic reinforcement (which we simply call ``reinforcement'' and ``meta-reinforcement'') in a multi-level network is the subject of the present study, which focuses on the dynamical creation and retrieval of a memorised pattern.
Specifically, we consider an idealised system constituted by a recurrent, feed-forward adaptive network that is traversed by a random walk (RW)~\cite{masuda_random_2017} and investigate how it can memorise and retrieve a pattern through the RW dynamics. If the nodes are interpreted as agents, the RW process models the communication between them; if they are interpreted as locations, the RW process models an active agent moving between different places. In both cases the memorisation and retrieval happen by changing the weights of the connections between different nodes so that  the RW is biased to choose paths that it has already explored: the memory is stored in the network itself. 

Biased RWs provide a standard model to investigate the mathematical properties of non-Markovian systems, for example their limit behaviours and phases~\cite{schutz_elephants_2004, holmes_senile_2007, agliari_true_2012, cressoni_alzheimer_2012, gut_elephant_2022}, including proofs of the walker being asymptotically confined to an attractor~\cite{donsker_number_1979, vlada_limic_attracting_2007, campos_nonstationary_2017, erhard_directed_2022}. 
To enable analytical treatment, such models feature particularly simple non-Markovian dynamics, like exponential or power-law reinforcement equations, resulting in weights that grow indefinitely and limiting their applicability to biophysical systems.

In our model, the weights of edges in the network grow according to a sigmoid function when the walker traverses them, resulting in a Hebbian reinforcement term that is asymptotically bounded by an upper value.
Meta-reinforcement increases the learning rate of the Hebbian dynamics and is 
stochastically distributed 
between edges belonging to local groups: this adds a second layer to the network, inducing a local coordination between nodes.
These assumptions reflect the presence of undefined interactions or unknown driving factors in the phenomena to model, e.g. unknown molecular mechanisms in astrocytes in a neuron-glial network or undetected, possibly off-line verbal communications between people in a social network. 
Following these scenarios, we further assume that reinforcement acts on a fast time-scale, and meta-reinforcement on a slower one.
Due to the complexity of our dynamics, we conduct our work mostly through numerical simulations.

Our results show that the addition of a second memorising component improves the capabilities of the model beyond those of regular reinforcement.
In particular, we obtain three regimes in the space of parameters, characterised by a component of the dynamics imposing itself on the others and controlling the outcome of the RW process.
In one of these regimes, meta-reinforcement is capable of driving the RW while reinforcement alone is ineffective;
moreover it allows for proper retrieval of the memorised path even after the Hebbian memory is completely erased.
Finally, we investigate the dependence of the dynamics on the size of the local groups subject to meta-reinforcement; we find that enlarging them broadens the range of parameters over which memory formation is successful, without affecting the qualitative behaviour of the system.

The manuscript is organised as follows.
In Sec.~\ref{sec:methods}, we define the model.
We describe the relevant observables, path entropy and path distance, in Sec.~\ref{sec:data_anal}.
In Sec.~\ref{sec:results}, we present the results of our simulations in terms of the formation of a memory engram and of its retrieval afterwards, as well as a comparison amongst various meta-reinforcement scales, testing various combinations of values for the parameters of our model.
Finally, in Sec.~\ref{sec:discussion} we analyse an asymptotic regime of the model and show how this can clarify some of the results obtained in Sec.~\ref{sec:results}.

\section{The model}
\label{sec:methods}

\subsection{Network structure}
\label{subsec:netstruct}
\begin{figure*}[!ht]
    \centering
    \includegraphics[width=\textwidth]{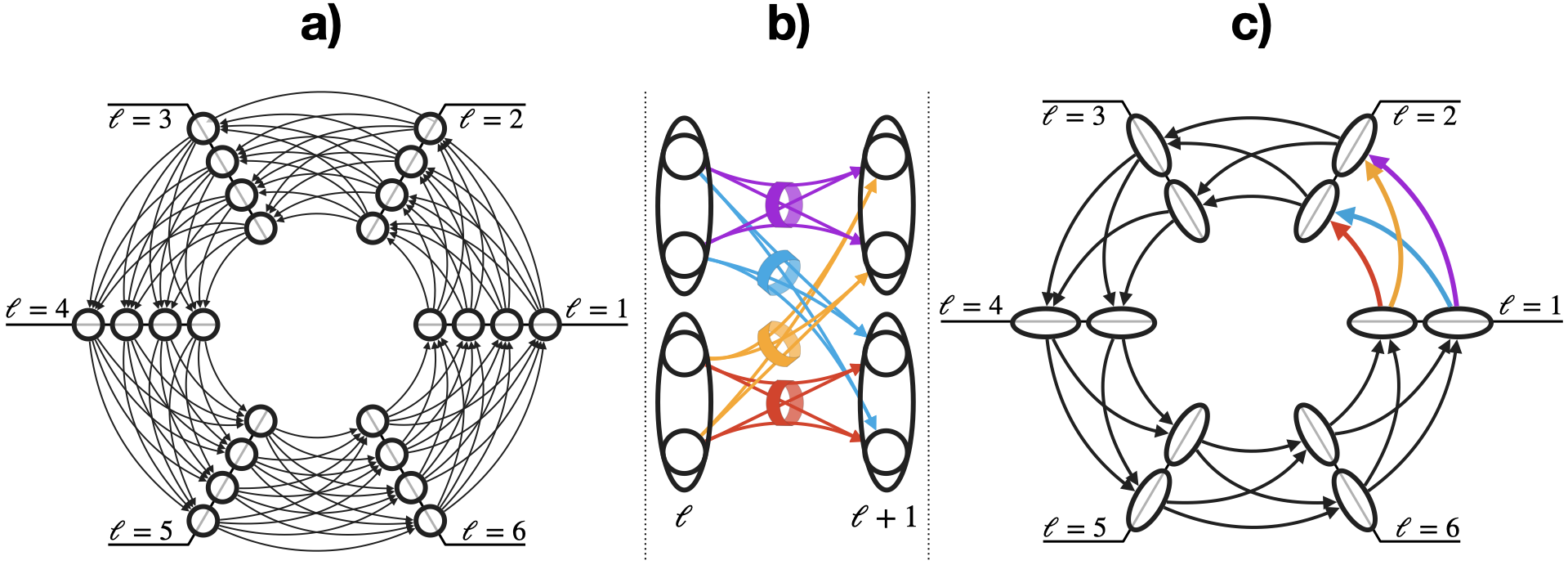}
    \caption{a) Example of cyclic feed-forward network with $6$ layers, labelled by $\ell$, with $4$ nodes each.
    b) The clustering of $M=2$ nodes from two consecutive layers of the network in panel~a), and the consequent grouping of the edges into a partition of $4$ edges groups (represented by the $4$ colours).
    Edges are distorted to ease visualisation of the grouping.
    c) The resulting network after edge grouping yields a lower scale (coarse-grained) network analogous (self-similar) to the one in panel~a);
    edge colours between layers $\ell =1$ and $\ell =2$ refer to the edge groups of panel~b).
    }
    \label{fig:structure_ensemble_cyclic}
\end{figure*}%
In this work we consider a \textit{cyclic feed-forward network} (CFFN).
This choice simplifies calculations while preserving the rich interplay between the two memory dynamics of the process.
As is standard for feed-forward networks, our network is divided in $L$ layers, labelled $\ell \in \left\{1,\ldots, L \right\}$, of $N$ nodes each, see Fig.~\ref{fig:structure_ensemble_cyclic}.
Each node in layer $\ell$ is connected to all nodes of layer $\ell+1$,
those of layer $\ell=L$ are connected to those of layer $\ell=1$, ensuring that the network has a recurrent structure (\textit{cyclic}).
All edges are directed from layer $\ell$ towards layer $\ell+1$.
The walker can only traverse them in this direction (\textit{feed-forward}).
Fig.~\ref{fig:structure_ensemble_cyclic}a) depicts the resulting network topology.
The specific CFFN we employ for the simulations presented in Sec.~\ref{sec:results} is composed of $L=20$ layers of $N=12$ nodes each. While this size has been chosen for numerical convenience, changing the number of layers or that of nodes does not modify the qualitative behaviour discussed in the rest of this manuscript.

To implement plasticity, the edges of the network are weighted proportionally to the number of times the walker has crossed them in the recent past.
We refer to this property as \textit{frecency} (a blend of frequency and recency) and through it we implement an Hebbian rule to bias the future movements of the walker, as described in Subsec.~\ref{subsec:reinf}.

To model a meta-plastic behaviour on a longer time-scale and across different nearby edges, we enrich the network with a second level of coordination between local groups of edges, which can be seen as a second system interacting with the CFFN.
First, we cluster the nodes in each layer $\ell$ in $M$-plets such that $M$ is an integer divisor of $N$, obtaining $\frac{N}{M}$ clusters or coarse-grained (CG) nodes per layer.
Then, we consider the $M^{2}$ edges going from the $M$ nodes of a cluster in layer $\ell$ to the $M$ of a cluster in layer $\ell+1$: these edges constitute one edge group $\group$ and share the meta-reinforcement contribution between them, proportionally to the weight of each one, see Sec.~\ref{subsec:meta-reinf}.
The grouping procedure is illustrated for $M=2$ in Fig.~\ref{fig:structure_ensemble_cyclic}b).
By construction, each edge belongs to one and only one group:
there are $\frac{N^{2}}{M^{2}}$ groups of $M^{2}$ edges between each layer of nodes.
This choice ensures that meta-reinforcement acts only locally, between edges connecting neighbouring nodes, and that the CG is self-similar to the full network, as shown in
Fig.~\ref{fig:structure_ensemble_cyclic}c).
In the rest of the manuscript we consider $M=2$. The effects of the CG scale $M$ on the dynamics of the system are explored in Sec.~\ref{subsec:CG_analysis}.

\subsection{Reinforcement dynamics}
\label{subsec:reinf}
To recover Hebbian-type memory dynamics, we employ an edge-reinforced random walk (ERRW) \cite{davis_reinforced_1990} and adapt it so that the memory may degrade, producing an effective short-term memory (STM) as explained below.

For a generic random walk on a network, the probability $\prob\left(i \rightarrow j\right)\equiv p_{ij}$ to go from node $i$ to node $j$ is given by the weight $w^{\edge}$ of the edge $\edge~=~i~\rightarrow~j$ connecting the two nodes, normalised over all edges leaving $i$:
\begin{equation}
\label{eq:RW}
    \prob\left(\edge\right)=\prob_{ij} = \frac{w^{\edge}}{\sum_{\edge' \in k^{+}\left(i\right)} w^{\edge'}},
\end{equation}
where $k^{+}\left(i\right)$ is the set of all out-going edges of node $i$.
In an ERRW, each edge weight is time-dependent:
it is increased (reinforced) each time the walker passes through it, and all edge weights decay over the same characteristic time-scale, see Eq.s~\eqref{eq:frecency}~-~\eqref{eq:sigmoid_logistic_function}.
The network becomes adaptive. 

Our implementation of an ERRW takes advantage of the CFFN topology described in Sec~\ref{subsec:netstruct}. 
A walker on an $L$-layered CFFN returns to the same layer after $L$ traversals, describing a loop around the network.
This allows one to fix the time unit of the simulation, $\Delta\loops =1$, as the time needed to perform $L$ traversals. 
This time unit makes parameters and observables independent on the number of layers in the network.

At the beginning of each loop, the frecencies of all edges (and consequently  their weights) are decreased. In each loop, edges that are traversed have their frecency increased by one. The frecency $s^\edge$ of edge $\edge$  thus evolves according to Eq.~\ref{eq:frecency}:
\begin{equation}
\label{eq:frecency}
    s^{\edge}_{\loops+1} = s^{\edge}_{\loops} e^{-\frac{1}{\tau}} + \mathds{I}{\left[\edge \in \walkerstep_{\loops}\right]}
    .
\end{equation}
where $\walkerstep_{\loops}=\left\{ \walkerstep_{\loops}^{\ell} \left.\right| \ell = 1,\dots,L\right\}$ is the sequence of $L$ edges, one per layer, traversed by the walker during loop $\loops$ and the term $\mathds{I}{\left[\edge \in \walkerstep_{\loops}\right]}$ increments by $1$ the frecency of those edges.
The frecency decrease is implemented through an exponential decay with dimensionless characteristic time $\tau$.
We take $s_\loops^{\edge}$ to be defined on $\mathbb{R}^{+}$ and set $s_{\loops=0}^{\edge} = 0$ for all edges.

\begin{figure}[t!]
    \centering
    \includegraphics[width=0.9\columnwidth]{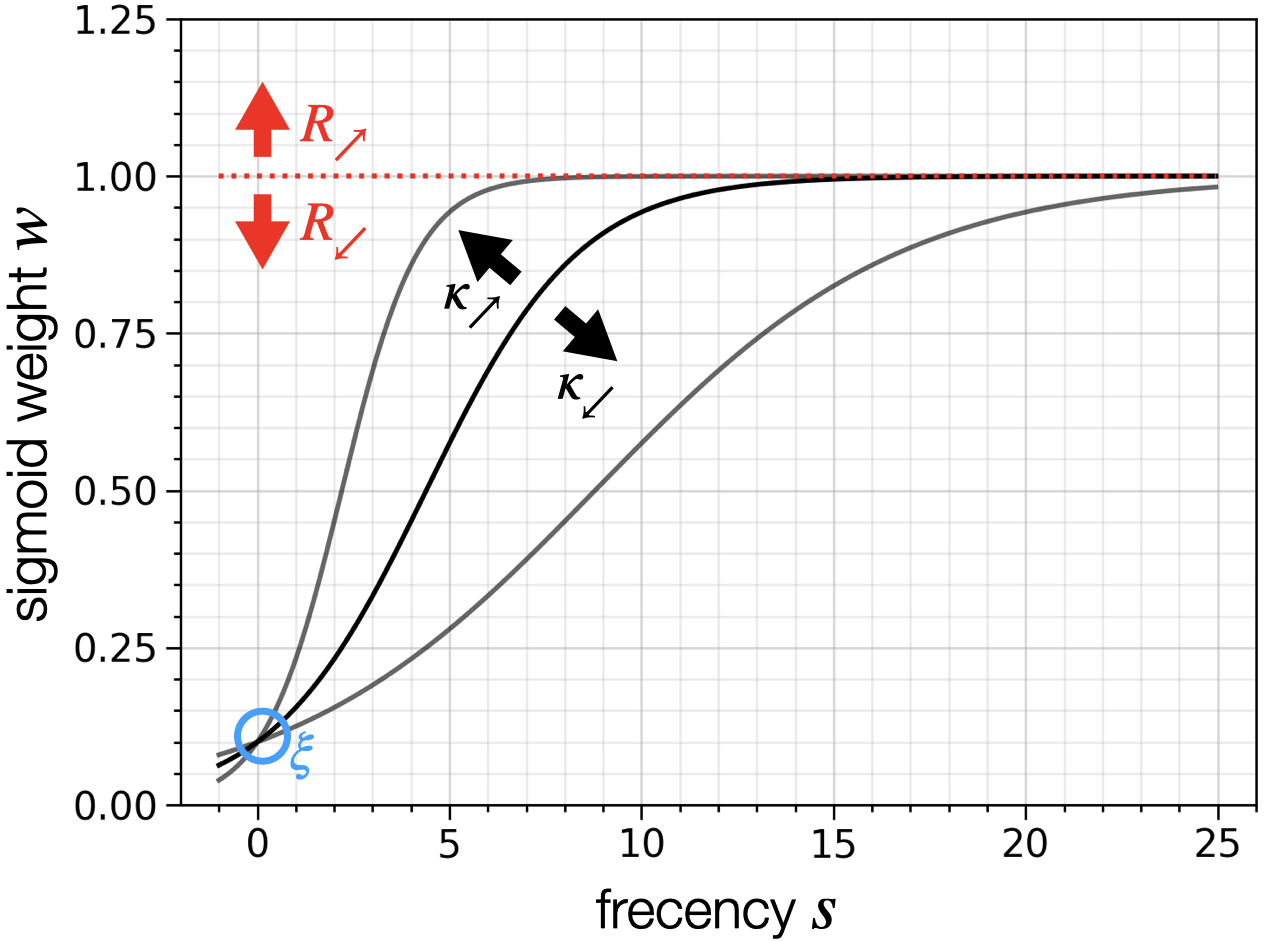}
    \caption{Examples of the dynamic on logistic functions $\sigma\left(s\right)$ versus frecency
    (black curves) for the regularised growth of edge weights, see Eq.~\ref{eq:sigmoid_logistic_function}, and the role of the parameters involved:
    $\xi$ ensures that in $s=0$ all functions have the same initial value regardless of the other parameters (see the blue circle).
    $R$ sets the upper asymptote for the function (red dotted line):
    lowering $R$ ($R_{\swarrow}$) shrinks the function while increasing it ($R_{\nearrow}$) stretches the function. 
    Finally, $\kappa$ controls how rapidly the function approaches its asymptotes, hence lowering $\kappa$ ($\kappa_{\swarrow}$) flattens the function while increasing it ($\kappa_{\nearrow}$) makes the function steeper (see text).}
    \label{fig:sigmoid_parameters} \end{figure}
    
To be compatible with any real and finite system, we bound the maximum value of the weights. We do so by modelling them as sigmoid functions of the frecency, so that for each edge $\edge$ the weight is $w^\edge_\loops = \sigma(s^\edge_\loops)$.
The model does not depend on the specific sigmoid function $\sigma$ adopted. For simplicity, we use the logistic function:
\begin{equation}
\label{eq:sigmoid_logistic_function}
   w^{\edge}_{\loops} = \sigma\left(s^{\edge}_{\loops}\right) = \frac{R}{1+\exp{\left(- \kappa s^{\edge}_{\loops} + \xi \right)}},
\end{equation}
which is analogous to the Hill equation employed in bio-chemistry to describe ligands binding as a function of their concentration~\cite{gesztelyi_hill_2012}.
Parameters  $\xi$, $R$, and $\kappa$ in Eq.~\eqref{eq:sigmoid_logistic_function} set respectively the minimum and maximum values of the weight, and the steepness of the curve. Their roles are reported graphically in Fig.~\ref{fig:sigmoid_parameters}.
We set $\xi=\ln{\left(NR-1\right)}$, such that the initial value of weights is always $w_0 = \nicefrac{R}{1+e^\xi} = \nicefrac{1}{N}$.

For ease of interpretation we adopt the quantity $\prob_{sup}\left(R\right)$, to represent the maximum probability for an edge to be traversed by a walker. This corresponds to the case in which the edge weight is $w_{sup}=R$, while all other edges leaving that node have weight $w_0$: 
\begin{equation}
    p_{sup}= \frac{w_{sup}}{w_{sup} + w_{0}\left(N -1 \right)} = \frac{R}{R+\frac{N-1}{N}}.
\end{equation}
$\prob_{sup}$ is never reached as
in general all weights $w_t>w_0$ $\forall\ t>0$, and $s_t$ cannot increase indefinitely, as explained in detail in Sec.~\ref{sec:discussion}.

Together, equations~\eqref{eq:RW}~--~\eqref{eq:sigmoid_logistic_function} describe an ERRW able to memorise how frequently each edge of the network was traversed in the near past.
This implements a STM, characterised by two independent time-scales: $\kappa$, in units of $\frac{1}{s}$, which sets the learning rate for the weight reinforcement, and $\tau$ for the decay time of $s$ itself, which leads to the decay of the weight to its base-value $w\left(0\right)=\frac{1}{N}$.

\subsection{Meta-reinforcement}
\label{subsec:meta-reinf}
The meta-reinforcement dynamics acts directly on the learning-rate $\kappa$ of the edges, making some connections more apt to be reinforced than others.
As explained before, this mimics a memory acting on long time-scales, without decay, and shared between neighbouring connections in a group.
No other details are considered, so the dynamics is assumed to be stochastic. 
This  dynamics is also bounded by a sigmoid $\sigma^{(\kappa)}$ (Eq.~\eqref{eq:meta_reinforcement_sigmoid}).
The learning rate $\kappa^{\edge}$ evolves as follows:
\begin{equation}
\label{eq:metareinforcement_sigmoid}
    \kappa^{\edge}_{\loops} = \kappa_{0} + \sigma^{(\kappa)}\left(y^{\edge}_{\loops}\right)
\end{equation}
where $\kappa_{0}$ is the initial learning rate and $y^{\edge}_{\loops}$ is a monotonically growing stochastic variable that models the random activation during loop $\loops$ of the edge-group, $\mathcal{G}$, to which edge $\edge_{\loops}$ belongs, see Sec.~\ref{subsec:netstruct}. 

The variable $y^{\edge}_\loops$ driving the time evolution of $\kappa$ is updated at each step according to a stochastic contribution that is split amongst the active edges within $\mathcal{G}$ according to their weights:
\begin{equation}
\label{eq:metareinforcement_dynamics}
    y^{\edge}_{\loops+1} =  y^{\edge}_{\loops} + c\frac{w^{\edge}_{\loops}}{\sum\limits_{\edge' \in \mathcal{G}} w^{\edge'}_{\loops}} \Theta{\left( w^{\edge}_{\loops} - w_{min} \right)} \chi^{\mathcal{G}}_{\loops},
\end{equation}
where $\Theta$ is the Heaviside theta function,  $c~\ge~0$ controls the rate of meta-reinforcement, and $\chi^{\group}_{\loops} \sim \mathcal{B}(\probmetareinf)$ is a stochastic variable following a Bernoulli process: $\chi^{\group}_{\loops} \sim \mathcal{B}(\probmetareinf)$ gives $1$ with probability $\probmetareinf$ and $0$ otherwise. 
We define the active edges as those for which $w_{\loops}^{\edge} \ge w_{min}$. Note that while edges that are not active are not able to experience the meta-reinforcement contribution, they are still counted in the denominator. Finally, we set 
$w_{min}=w\left(s=e^{-\frac{5}{\tau}}\right)$ so that edges weaker than an edge traversed only once 5 loops in the past are ignored.

The $\sigma^{(\kappa)}$ function of Eq.~\eqref{eq:metareinforcement_sigmoid} is an edge-independent shifted logistic function with constant parameters:
\begin{equation}
\label{eq:meta_reinforcement_sigmoid}
    \sigma^{(\kappa)}\left(y\right) = \frac{2.5}{1+\exp{\left(- y\right)}} - 1.25.
\end{equation}
This form ensures that $\sigma^{(\kappa)}(0) = 0$ and $\sigma^{(\kappa)}(\infty) = 1.25$, so that $\kappa_{\loops}^{\edge} \leq 2.25$ given $\kappa_0=1$.
Furthermore, the $-1.25$ shift  makes meta-reinforcement most responsive for small values of $y$.
Finally, we normalise the steepness of $\sigma^{(\kappa)}$ to $1$ for simplicity.

Through Eq.s~\eqref{eq:metareinforcement_sigmoid}~--~\eqref{eq:metareinforcement_dynamics}, the meta-reinforcement dynamics adds a long-term memory built on top of the STM introduced in Sec.~\ref{subsec:reinf}.
This memory form never fades:
it is unaffected by the decay in frecency from Eq.~\eqref{eq:frecency} and we do not consider any decay mechanism for $y$ in Eq.~\eqref{eq:metareinforcement_dynamics} as we are assuming that it would happen over a much larger time-scale compared to any other involved in the model and can hence be neglected.
The time-scale for the meta-reinforcement dynamics is controlled by both the meta-reinforcement coefficient $c$ and the probability $\probmetareinf$ for the stochastic variable in Eq.~\eqref{eq:metareinforcement_dynamics}.
Having the contribution per unit time-step be split amongst $M^{2}$ edges, proportionally to the relative weights, further slows the dynamics and hence increases its time-scale.

\section{Observables}
\label{sec:data_anal}
We analyse the capability of the system to memorise through two quantities: a) the entropy rate of the biased RW, and b) a layer-wise distance between different paths. Both are defined as functions of time, measured in loops $\loops$ on the CFFN, to describe memory as it emerges. 

\subsection{Entropy rate}
\label{par:entropy_rate}
The information content of a stochastic variable can be described by the Shannon entropy $H(X) = -\int p(x) \log(p(x))dx$, which gives an estimate of the number of ``bits'' required to describe its possible output. In the case of a stochastic process in which (possibly correlated) random variables are sampled one after the other, the Shannon entropy grows with the number of variables $n$, and it is thus simpler to study its rate of growth, called the \emph{entropy rate}~\cite{cover_entropy_2005}. Mathematically, this is defined as the limit
\begin{equation}
\label{eq:entropy_rate_limit_definition}
    H\left( \mathcal{X} \right) = \lim_{n\to\infty}\frac{1}{n}H\left(X_{1},X_{2},\dots,X_{n}\right)
\end{equation}
where $\left\{ X_{k} \right\}$ is a stochastic process, and $H$ is the joint Shannon entropy of the set of $n$ random variables $X_a$.
Stochastic processes that require less information to be described  have a low entropy rate, while processes that require more information  have a high entropy rate. This makes such a measure particularly useful for studying an ERRW, as one can expect that the entropy rate of a walker that has memorised a single path will be particularly low, while it will be maximal for a random walk on a uniformly-weighted network.

As our system evolves with time, we defined a time-dependent entropy rate $H_\loops$, able to capture the transition from a non-memorising to a memorising system. To do so we leverage the fact that the entropy rate of a Markov process can be computed simply from the transition matrix of the process, $P_{ab}$, and the stationary occupation vector $\mathbf{p}^{*}_a$ of all states $a$ as in Eq.~\eqref{eq:entropy_rate}~\cite{cover_entropy_2005}: 
\begin{equation}
\label{eq:entropy_rate}
    H=-\frac{1}{\mathcal{N}}\sum_{ab} \mathbf{p}^{*}_{a} P_{ab} \ln{\left(P_{ab}\right)},
\end{equation}
where we added a normalisation factor $\mathcal{N}$  set so that $H=1$ for a transition matrix in which all non-zero terms are equal.

We interpret the $NL\times NL$ weighted adjacency matrix $A_{ab}$ of the full network at time $\loops$,   Eq~\eqref{eq:adj_matrix_CFF}, as the transition matrix $P_{ab}$ of a Markov process.
Taking advantage of the CFF topology of the network, $A_{ab}$ can be written as a block matrix: 
\begin{widetext}
\begin{equation}
\label{eq:adj_matrix_CFF}
    A_{ab} = 
    \bordermatrix{ \bsfrac{\ell'}{\ell} & 1 & 2 & 3 &  \dots & L-2 & L-1 & L \cr
      1 & \emptyset & T^{1,2} & \emptyset & \dots & \emptyset & \emptyset & \emptyset \cr
      2 &\emptyset & \emptyset & T^{2,3} & \dots & \emptyset & \emptyset & \emptyset \cr
      3 & \emptyset &\emptyset & \emptyset & \dots & \emptyset & \emptyset & \emptyset \cr
      \vdots & & & & \ddots & & & \cr
      L-2 & \emptyset & \emptyset & \emptyset & \dots & \emptyset & T^{L-2,L-1} & \emptyset \cr
      L-1 & \emptyset & \emptyset & \emptyset & \dots & \emptyset & \emptyset & T^{L-1,L} \cr
      L & T^{L,1} & \emptyset & \emptyset & \dots & \emptyset & \emptyset & \emptyset }
\end{equation}
\end{widetext}
where blocks are labelled by layer, $1$ to $L$, and the sub-matrices $\tensor{T}{^{\ell}_{i}^{\ell +1}_{j}}$ are the $N\times N$ transition matrices from one layer to the next, expressed in terms of the probabilities $p_{ij}$ introduced in Eq.~\eqref{eq:RW}:
\begin{equation*}
    \tensor{T}{^{\ell}_{i}^{\ell +1}_{j}} = 
    \bordermatrix{
    \bsfrac{j}{i} & 1 & 2 & \dots & N \cr
    1 & p_{11} & p_{12} & \dots & p_{1N} \cr
    2 & p_{21} & p_{22} & \dots & p_{2N} \cr
    \vdots & & & \ddots & \cr
    N & p_{N1} & p_{N2} & \dots & p_{NN}
    }
\end{equation*}
Note that with the CFF topology, layer $L$ is connected to layer $1$.

Since $A_{ab}$ is an irreducible stochastic matrix, one can find the stationary occupation probability distribution $\mathbf{p}^{*}$ associated to it through the Perron-Frobenius theorem:
$\mathbf{p}^{*}$ is the left-eigenvector associated to the largest real eigenvalue of $A_{ab}$~\cite{s_u_pillai_perron-frobenius_2005}. This calculation is repeated at each step $\loops$ and plugged into Eq.~\ref{eq:entropy_rate} to obtain $H_\loops$.
\subsection{Distance between paths}
To further analyse our results we need to compare the distance between a path, i.e. a sequence of nodes $\Gamma = \left(n^{\ell=1}, \dots, n^{\ell=L}\right)$  and a meta-reinforced, CG path, i.e. a sequence of CG nodes $\Gamma_{CG} = \left(\{n\}_{CG}^{\ell=1}, \dots, \{n\}_{CG}^{\ell=L}\right)$, where $\{n\}_{CG}^{\ell}$ represents a set of nodes corresponding to one of the CG nodes  depicted in Fig.~\ref{fig:structure_ensemble_cyclic}c). Leveraging the CFFN structure of our network, we can define a distance between any two paths $\Gamma_{1}$ and $\Gamma_{2}$ as a rescaled Hamming distance:
\begin{equation}
\label{eq:Hamming_paths_exact}
	d\left(\Gamma_1,\Gamma_2\right) = \frac{N}{N-1}\left(1- \frac{1}{L} \sum_{\ell=1}^{L} \mathds{I}{\left[\tensor{n}{_{1}^{\ell}} = \tensor{n}{_{2}^{\ell}}\right]}\right),
\end{equation}
where the sum counts the number of nodes of $\Gamma_1$ coinciding with those of $\Gamma_2$ and the factor $\frac{N}{N-1}$ ensures that on average $\langle d\left(\Gamma_1,\Gamma_2\right)\rangle = 1$ when either path is composed of nodes picked randomly from a uniform distribution, corresponding to a RW without any memory. To compute the distance between a path $\Gamma$ and a CG path $\Gamma_{CG}$, we modified Eq.~\ref{eq:Hamming_paths_exact} as follows:
\begin{equation}
\label{eq:Hamming_paths}
	d\left(\Gamma,\Gamma_{CG}\right) = \frac{N}{N-M} \left(1- \frac{1}{L} \sum_{\ell=1}^{L} \mathds{I}{\left[n^{\ell} \in \{n\}_{CG}^{\ell}\right]}\right),
\end{equation}
where the sum $\sum_{\ell=1}^{L} \mathds{I}{\left[n^{\ell} \in \{n\}_{CG}^{\ell}\right]}$ counts the number of layers in which the node in $\Gamma$ lies within a CG node in $\Gamma_{CG}$, and the rescaling factor takes into account that there are $M$ nodes in each layer for which $\mathds{I}{\left[n^{\ell} \in \{n\}_{CG}^{\ell}\right]}=1$. For $M=1$, $d\left(\Gamma,\Gamma_{CG}\right) = d\left(\Gamma_1,\Gamma_2\right)$. Note that while Hamming distances are usually defined between $0$ and $1$, this is not true in our case due to the rescaling.
While path distances are intrinsically noisier compared to the entropy rate, the two quantities are complementary. The latter gives us a measure of the memorisation of the system, while the former can be used to measure how precisely a given path has been memorised.

\begin{figure*}[ht!]
    \centering
    \includegraphics[width=0.9\textwidth]{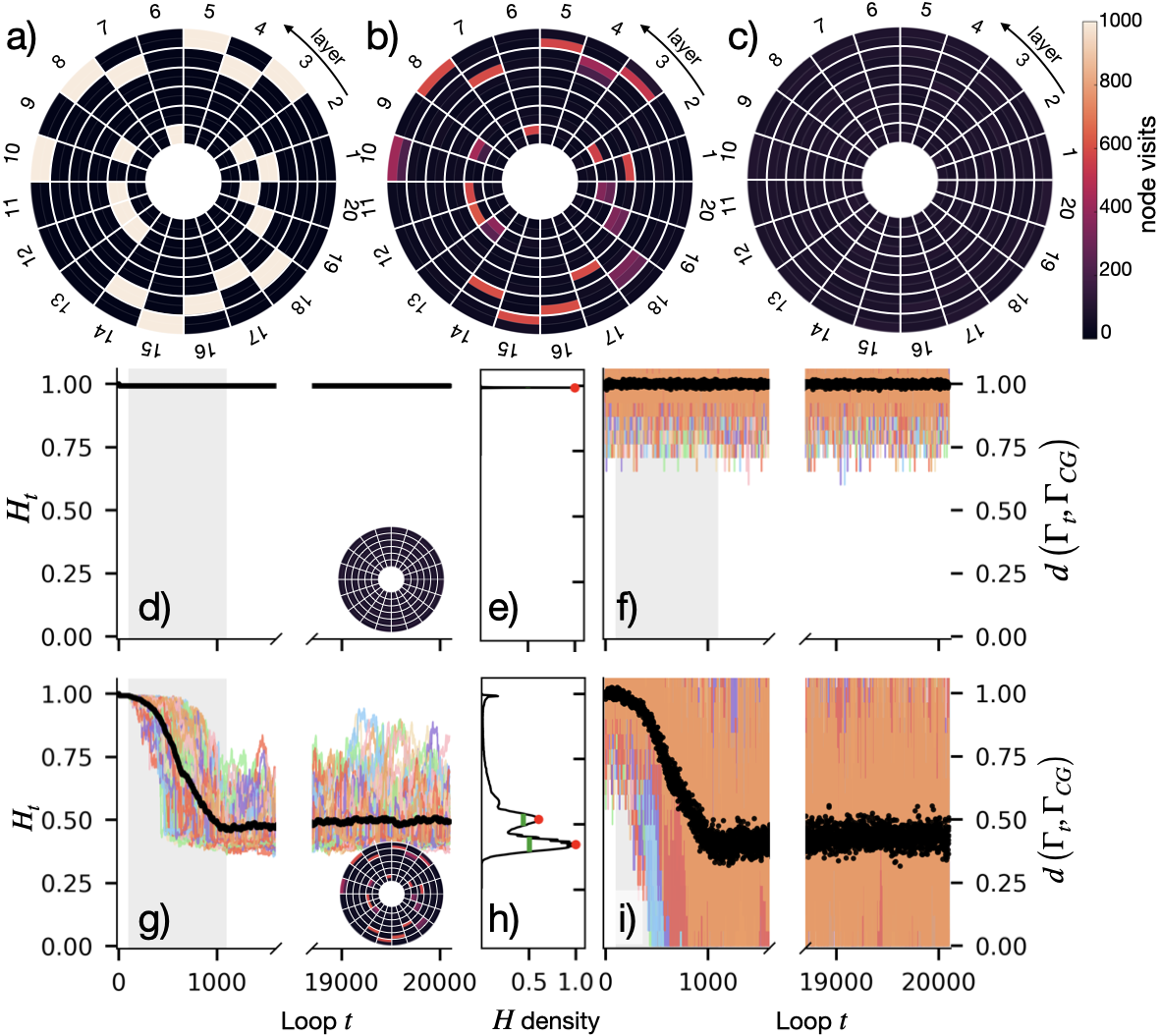}
    \caption{
    a) A coarse-grained path $\Gamma_{CG}$ to be memorised on an  network with CG level $M=2$. 
    b) - c) Heatmaps of the visits (see colour-bar on the right) of different nodes during the final 1000 steps for an individual realisation of: b) a memorising process ($\prob_{sup}=0.95$, $c=2$, $\tau=8$) with $M=2$; and c) a process that does not display memory ($\prob_{sup}=0.95$, $c=0$, $\tau=8$).
    White lines in a), b) and c) divide the network in layers and $M=2$-plets of nodes, see Fig.~\ref{fig:structure_ensemble_cyclic}.
    The time evolution of the normalised $H_\loops$ and the distance from the target path $d\left(\Gamma_{\loops},\Gamma_{CG}\right)$ for the systems depicted in b) and c) are reported in panels g)-h) and d)-f) respectively. Coloured lines represent $100$ independent realisations, black bullets their average at loop $\loops$, and the grey shaded area the period $T_2$ of meta-reinforcement.
    The histograms of $H_{\loops}$ are reported in panels e) and h). 
    Peaks are indicated in red, full-widths-at-half-maximum in green (visible in h)); for clarity the histogram is normalised to the highest peak.}
    \label{fig:example_of_process}
\end{figure*}

\section{Numerical results}
\label{sec:results}
We perform numerical simulations of a biased random walker on a CFFN, as described in sec.~\ref{sec:methods}.
The simulations are structured so that the dynamics goes through three stages. For the first $T_1$ steps (loops) the system follows a normal Hebbian dynamics, without meta-reinforcement. This allows us to compare its effect with meta-reinforcement, which is activated on top of Hebbian reinforcement during the second stage, lasting $T_2$ steps.
Finally, the system continues to evolve following a normal Hebbian dynamics, without meta-reinforcement, for $T_3$ steps.

For simplicity, we consider an idealised case in which only one sequence of edge groups, corresponding to a CG path $\Gamma_{CG}$, gets meta-reinforced during $T_2$ as shown in Fig.~\ref{fig:example_of_process}a).
We identify $\Gamma_{CG}$ with the set of consecutive CG nodes that contains the path chosen by the walker during the first loop, $\Gamma_1$, and connect the $L-$th and the first node so that $\Gamma_1$ and $\Gamma_{CG}$ are closed paths.

We characterise the formation of memories (paths) by computing the entropy rate $H_\loops$ as this decreases with the emergence of preferred paths as shown in Fig.~\ref{fig:example_of_process}g), and the interplay between reinforcement and meta-reinforcement by computing the distance $d\left(\Gamma_{t},\Gamma_{CG}\right)$ between $\Gamma_{CG}$ and a path memorised at times $t>>T_1+T_2$.
For the reference systems without meta-reinforcement ($c=0$), we compute instead the distance $d\left(\Gamma_{t},\Gamma_{1}\right)$ from the exact path $\Gamma_{1}$.
This protocol further allows us to test memory storing via meta-reinforcement alone.
In this case we set all the weights to their initial values $w_{0}$ after $T_1 + T_2$ steps, thus erasing any Hebbian memory, and let the RW proceed for $T_3$ steps, checking whether a path $\Gamma_t$, $t>>T_1+T_2$ is close to $\Gamma_{CG}$, meaning that the Hebbian dynamics in $T_3$ is able to retrieve a path stored through meta-reinforcement.

We investigate the following sets of parameters:
\begin{itemize}
    \item $\tau~\in~\{ 4, 2^{\nicefrac{5}{2}},8, 2^{\nicefrac{7}{2}},16,32,64\}$,
    \item $c~\in~\{ 0, 0.25, 0.5, 1, 2, 4, 8\}$,
    \item $p_{sup}~\in~\{0.85, 0.95, 0.999\}$.
\end{itemize}
We set $T_1~=~100$, $T_2~=~1000$, and $T_3~=~19000$.  These times allow  to observe both reinforcement and meta-reinforcement effects, and their interplay.
Their precise values do not affect the qualitative results obtained in this study, as shown in Supplementary material.
$T_3$ in particular is sufficiently long to guarantee that the dynamics reaches a steady state for each combination of parameters.
We perform $100$ independent realisations for each combination of parameters $p_{sup}$, $\tau$, and $c$ and average all observables over them.
Two example trajectories, one for a system showing memory ($\prob_{sup}~=~0.95$, $c~=~2$, $\tau~=8~$) and the other for a system that does not ($\prob_{sup}~=~0.95$, $c~=~0$, $\tau~=~8$), are reported in Fig.~\ref{fig:example_of_process}b) and~c) respectively.
To remove fluctuations, we compare the memorisation ability of different systems by considering the value of $H_\loops$ and $d\left(\Gamma_t,\Gamma_{CG}\right)$ averaged over 100 realisations and over the last 1000 steps of the simulation, when they have reached a stationary value. We indicate this double average with the symbol $\savg{\cdot}$.

\subsection{Memory formation}
To observe whether a system is able to memorise, we compute the entropy rate $H_t$ and produce a histogram of its possible values.
An illustrative example is reported in Fig.~\ref{fig:example_of_process}d+e) for $p_{sup}~=~0.95$, $c~=~0$, $\tau~=~8$ (no memory), g+h) for $p_{sup}~=~0.95$, $c~=~4$, $\tau~=~8$ (memorising).
One can readily see that a memorising system shows a sharp decrease in entropy rate during the meta-reinforcement window $T_2$ (in grey), while the other one remains constant after an initial small decrease due to reinforcement.
Both systems reach a stable state in $\left\langle H_\loops \right\rangle$ (averaged over realisations, black bullets),  also captured by the histograms.
The memorising system shows well-separated peaks:
one at $H_\loops~\gtrsim~0.98$, corresponding to the non-meta-reinforced process happening at the beginning of the simulation, and others at or below $H_\loops~\sim~0.6$.
The low average distance $\langle d\left(\Gamma_{\loops},\Gamma_{CG}\right) \rangle$ reported in Fig.~\ref{fig:example_of_process}i) (black bullets) indicates that the lower entropy state corresponds to the walker traversing the meta-reinforced path $\Gamma_{CG}$.

\begin{figure*}[ht!]
    \centering
    \includegraphics[width=0.9\textwidth]{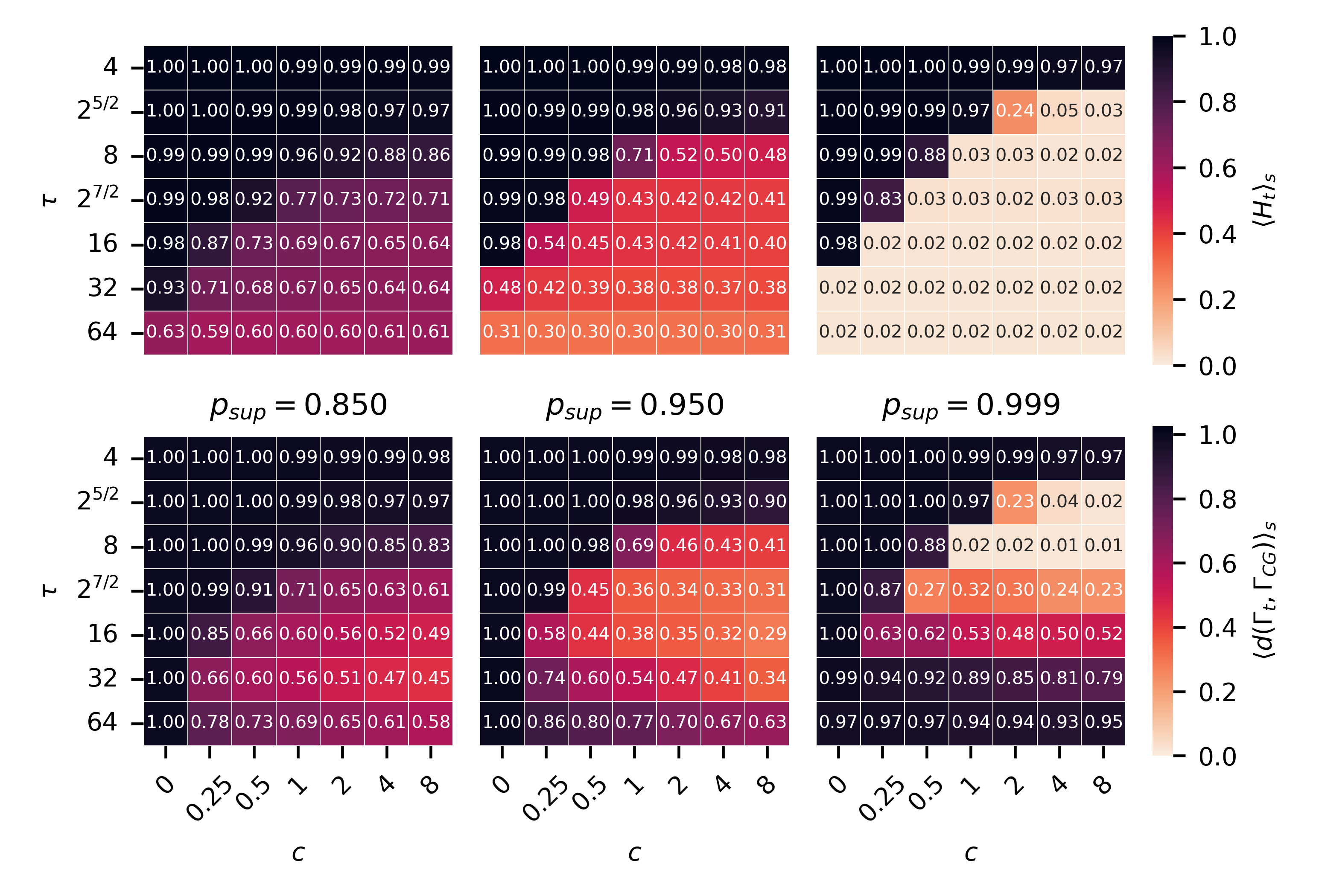}
	\caption{Heatmaps reporting stable states of the average entropy rate $\savg{H_\loops\left(\loops\right)}$ (top) and distance from the coarse-grained target path $\savg{d\left(\Gamma_{\loops},\Gamma_{CG}\right)}$ (bottom) as a function of decay time $\tau$ and meta-reinforcement contribution $c$ for different values of maximal probability $p_{sup} \in \{ 0.85, 0.95, 0.999\}$. The columns corresponding to $c=0$ in bottom heatmaps are not indicative of any system behaviour, as in absence of meta-reinforcement the distance from target path $\Gamma_{CG}$ is uninformative.}
    \label{fig:heatmap_steady_k0_100_T_100}
\end{figure*}
The behaviour of the entropy rate and distance reported in Fig.~\ref{fig:example_of_process} allow one to identify a steady state in the dynamics.
In all cases, this is reached well before the last $1000$ steps in the run, and in general shortly after $T_1+T_2$.
The results, reported in Fig.~\ref{fig:heatmap_steady_k0_100_T_100}, show several interesting effects of meta-reinforcement. 
First, meta-reinforcement can drive the formation of memory when reinforcement alone fails.
This is evident from the behaviour of $\savg{H_\loops}$ and $\savg{d\left(\Gamma_t, \Gamma_{CG}\right)}$ for $2^{\nicefrac{5}{2}} \leq \tau \leq 16$, where one can observe that increasing $c$ significantly lowers both the entropy rate and the distance.
This transition is particularly sharp for $p_{sup}=0.999,\, \tau=8$, with both observables decreasing by more than $95\%$ for $c \geq 1$:
the system goes from a situation like the one reported in Fig.~\ref{fig:example_of_process}d+f) to one like that reported in Fig.~\ref{fig:example_of_process}g+i).
Second, increasing reinforcement (through the increase of $\tau$) leads to processes that memorise a random path even at high levels of meta-reinforcement, as can be seen from the $\tau = 32, 64$ rows in the heatmaps of $\savg{d\left(\Gamma_{\loops}, \Gamma_{CG}\right)}$ in Fig.~\ref{fig:heatmap_steady_k0_100_T_100}.
Again, this is particularly evident for the case $p_{sup}=0.999$, where dynamics with $\tau = 64$, while having very low values of $\savg{H_\loops}$, display values of $\savg{d\left(\Gamma_{\loops},\Gamma_{CG}\right)}$ that are very close to those of a non-memorising case.
Considering the parameter space explored in our work, the optimal combination is found for $c=8$, $\tau = 2^{\nicefrac{5}{2}},8$.
For $\tau=4$, no value of $c$ could lead the system to memorisation.
Varying $T_1$ does not change these results in any meaningful way, as shown in the Supplementary Material.

\begin{figure}[t]
    \centering\includegraphics[width=0.9\columnwidth]{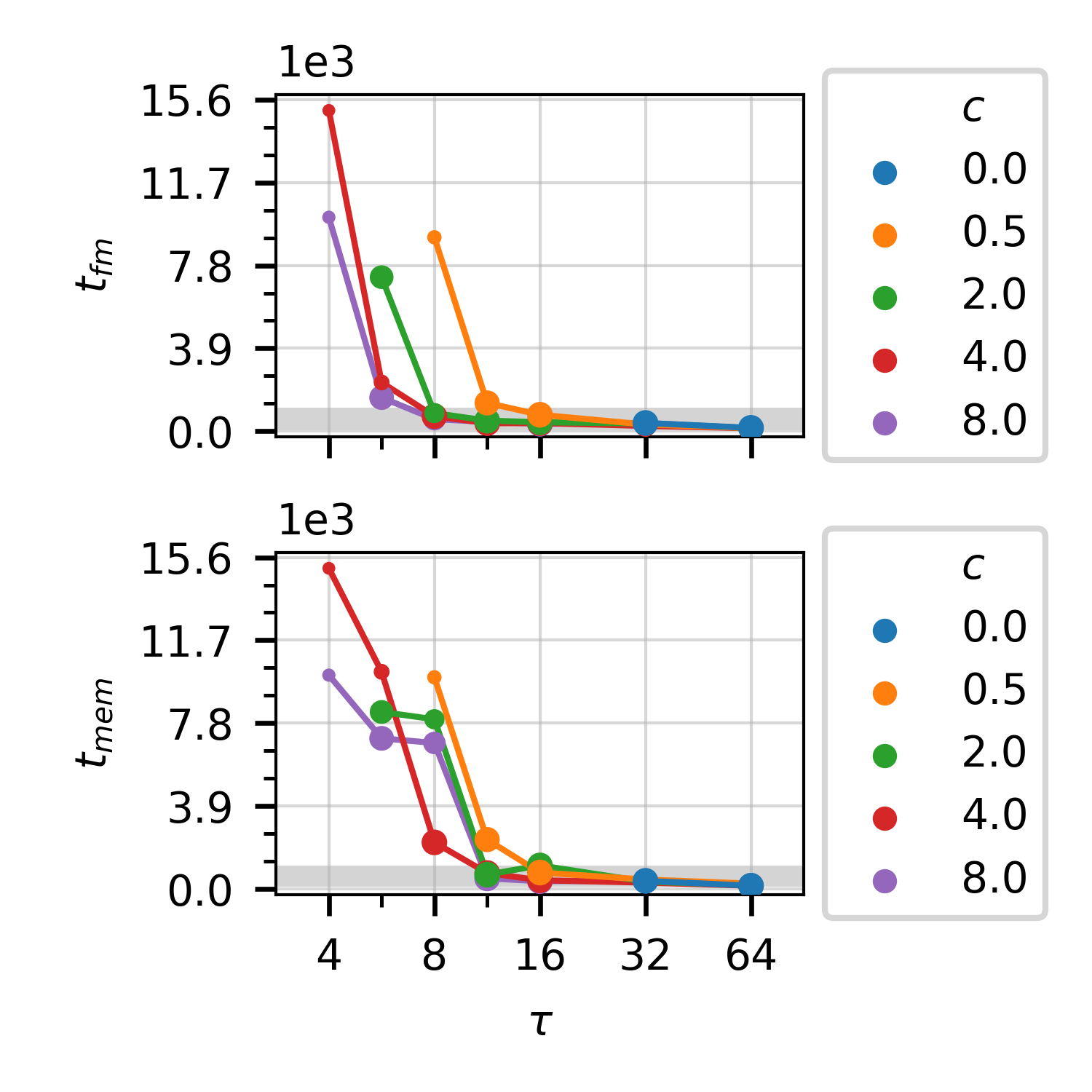}
    \caption{Average time, $\loops_{fm}$, to first reach a memorising state with $H_\loops<H^*$ (top) and memorisation time, $\loops_{mem}$, (bottom) as functions of the decay times $\tau$ (in $\log_{2}$-scale).
    Different curves correspond to different values of the meta-reinforcement contribution $c$ (colours) for $\prob_{sup}=0.999$.
    Values are averaged over runs that did reach a state below $H^{*}$:
    bullet sizes reflect the proportion of such runs.
    The grey shaded area indicates the meta-reinforcement window from $\loops=T_{1}$ to $\loops=T_{1}+T_{2}$.}
    \label{fig:memorisation_times_k0_100_T_100}
\end{figure}

The next question to consider is whether meta-reinforcement also affects the time needed to memorise a path (or a subset of paths).
To introduce such a time, we associate a state of the system to each peak of the histogram of $H_t$. We say that the system is in a given state whenever $H_t$ lies between the lower and upper values, $H^-$ and $H^+$, that delimit the full-width-at-half-maximum of the peak (FWHM, in green in Fig.~\ref{fig:example_of_process}). The time to reach a state is defined as the first time for which $H_t<H^+$ for that state, averaged  over 100 realisations.

We consider a state to entail memory formation if it lies below a memorisation threshold $H^*$ defined as the largest value of $\savg{H_t}$ for which $\savg{d(\Gamma_t,\Gamma_{CG})} \leq 0.5$.
From Fig.~\ref{fig:heatmap_steady_k0_100_T_100}, we see that this happens for $\prob_{sup}=0.85$, $c=8$,  $\tau=16$ where $\savg{d}=0.49$ for $\savg{H_t}=0.64$, so $H^{*}=0.64$.

We focus on two specific states: the very first memorising state ($H<H^*$) to be reached, $H_{fm}$, and the most populated memory state, $H_{mem}$. Their corresponding times, $\loops_{fm}$ and $\loops_{mem}$, indicate the minimum time to form a memory, and the time required to form a stable memory, respectively.
Comparing $\loops_{fm}$ and $\loops_{mem}$ we can address the presence of possible ``meta-stable states'' of entropy rate $H$ larger than $H_{mem}$ and below the memorisation threshold $H^{*}$, see e.g. Fig.~\ref{fig:example_of_process}h).
Naturally, if there is only one entropy rate state below $H^{*}$ then $\loops_{mem}$ and $\loops_{fm}$ will coincide.
Since not all runs reach a memory state, we compute the average values of $\loops_{fm}$ and $\loops_{mem}$ only over those runs in which memorisation did happen.

Fig.~\ref{fig:memorisation_times_k0_100_T_100} depicts the results obtained as a function of the decay time $\tau$ for  $c=0, 0.5, 2, 4, 8$ (colours), 
$\prob_{sup}=0.999$.
Bullet sizes reflect the fraction of memorising runs.
The average number of steps to reach the memorising states, $\loops_{fm}$ (top) and $\loops_{mem}$ (bottom), depend on both $\tau$ and $c$.
Both observables reach low steady values for $\tau=32$ and $\tau=64$ with no significant difference as $c$ changes:
this memory is unrelated to meta-reinforcement, as highlighted also for Fig.~\ref{fig:heatmap_steady_k0_100_T_100}.
For $\tau<32$, $\loops_{fm}$ decreases monotonically with increasing $c$.
Increasing $\tau$ reduces the impact of weight decay while increasing $c$ reduces the time-scale of meta-reinforcement, both factors leading to a more rapid memorisation.
On the other hand, $\loops_{mem}$ shows the same decreasing behaviour in $\tau$, but a more intricate one in $c$.
For example, for $c=4$ $\loops_{mem}$ is larger than for $c=2$ when $\tau=2^{\nicefrac{7}{2}}$ but substantially lower than for $c=8$ when $\tau=8$.
Comparing with the monotonic behaviour of $\loops_{fm}$, this behaviour seems to indicate that certain combination of parameters can lead to the emergence of meta-stable states.
In the Supplementary Material we report results for all values of $\prob_{sup}$ and $c$ considered in this work.

\subsection{Memory retrieval}
\begin{figure*}[ht!]
    \centering
    \includegraphics[width=0.8\textwidth]{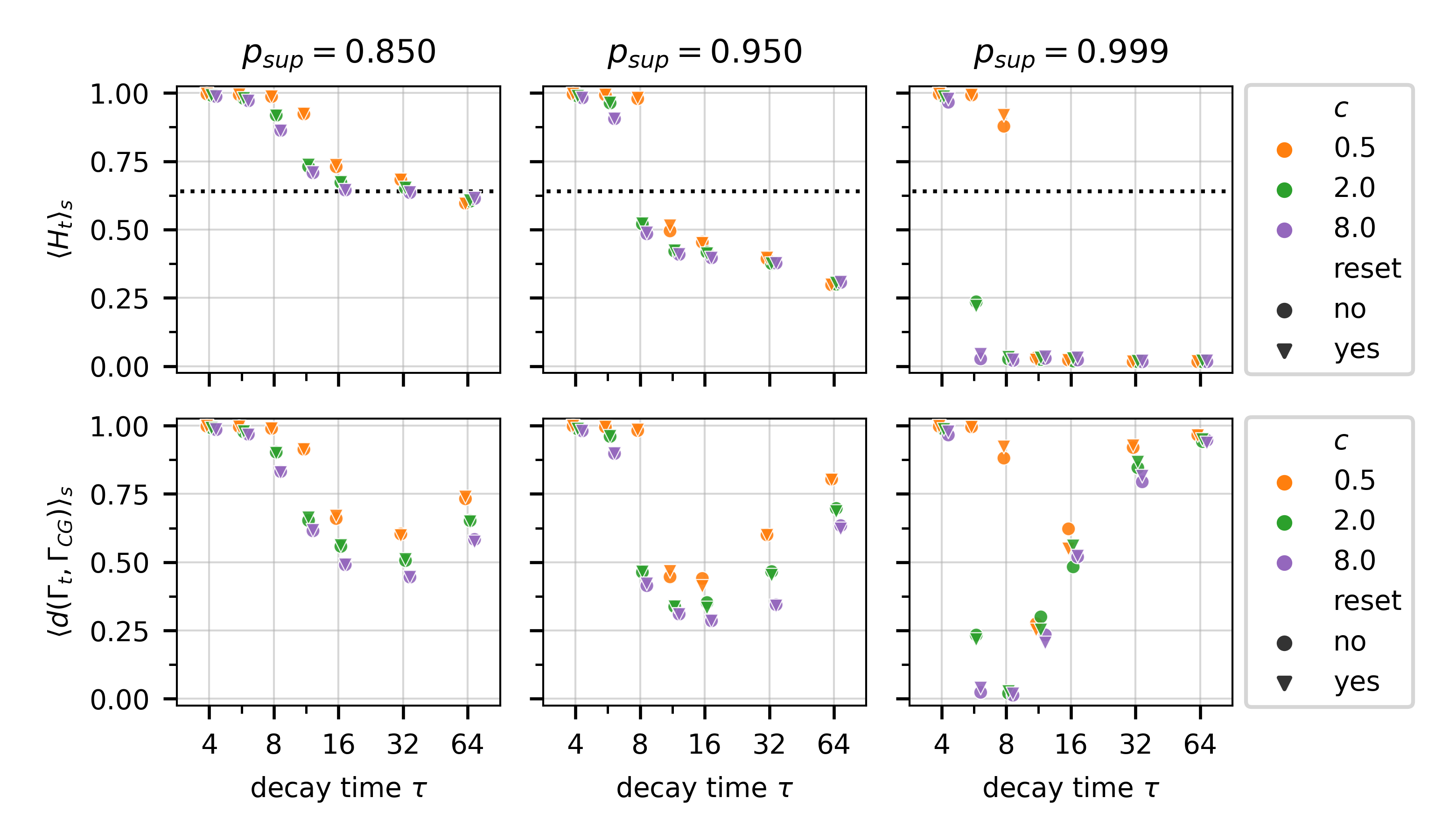}
    \caption{Memory retrieval analysis through entropy rate (above) and distance (below) from the steady-state distributions as functions of the decay times $\tau$ ($\log_{2}$ scale), the meta-reinforcement coefficient $c$ (colours) for different values of $p_{sup} \in \{ 0.85, 0.95, 0.999\}$.
    Symbols are slightly displaced along the $x$-axis to ease visualisation and comparison.
    Bullet shape reflects when the steady-state is computed:
    before (circular bullets) or after (triangular bullets) the reset.
    The dotted line in the top panels represents the $H^*=0.64$ entropy rate threshold associated to memorisation.}
    \label{fig:recall_scatters_k0_100_T_100}
\end{figure*}
We now check whether a meta-reinforced path can be retrieved when all weights have been reset to zero, and the only difference between edges lies in the learning rates $\kappa^\edge$, which are steeper for the meta-reinforced ones.
This property is essential in systems, like the brain, that need to store new memories while maintaining access to old ones.
To test this, we perform a series of runs in which at the end of the meta-reinforcement window $\loops = T_1+T_2$, all variables $s^\edge_t$ are reset to zero, and all weights $w_t^\edge$ to $w_{0}$.
The RW then proceeds following the standard Hebbian dynamics described in Sec.~\ref{subsec:reinf} for $T_3$ steps.
Since this reset erases any and all short-term memory from the network, we do not consider the $c=0$ case where meta-reinforcement is not present and any emerging memory is short-term.
As before, we measure the entropy rate $H_t$ and the distance from the meta-reinforced path $d\left( \Gamma_\loops, \Gamma_{CG} \right)$ as function of $\loops$, averaged over $100$ independent runs, and consider their steady state values averaged over the last 1000 steps.
The results, reported in Fig.~\ref{fig:recall_scatters_k0_100_T_100}, show that when the weights are reset, meta-reinforcement is still able to recall the correct path with a precision that is overall comparable to that obtained in Fig.~\ref{fig:heatmap_steady_k0_100_T_100} without the reset.
In the Supplementary Material we report results for all values of $\prob_{sup}$ and $c$ considered in this work, showing no qualitative change compared to what is reported here.

\subsection{Effect of meta-reinforcement scale}
\label{subsec:CG_analysis}
Finally, it is interesting to investigate how changing the size of the CG nodes, $M$, affects the results. Here, we fix $p_{sup}=0.999$ and consider $M=1$ (no CG), $2$, $3$, $4$, and $6$ (results for $p_{sup}=0.95$ are reported in the Supplementary Material).
\begin{figure*}[ht!]
    \centering
    \includegraphics[width=\textwidth]{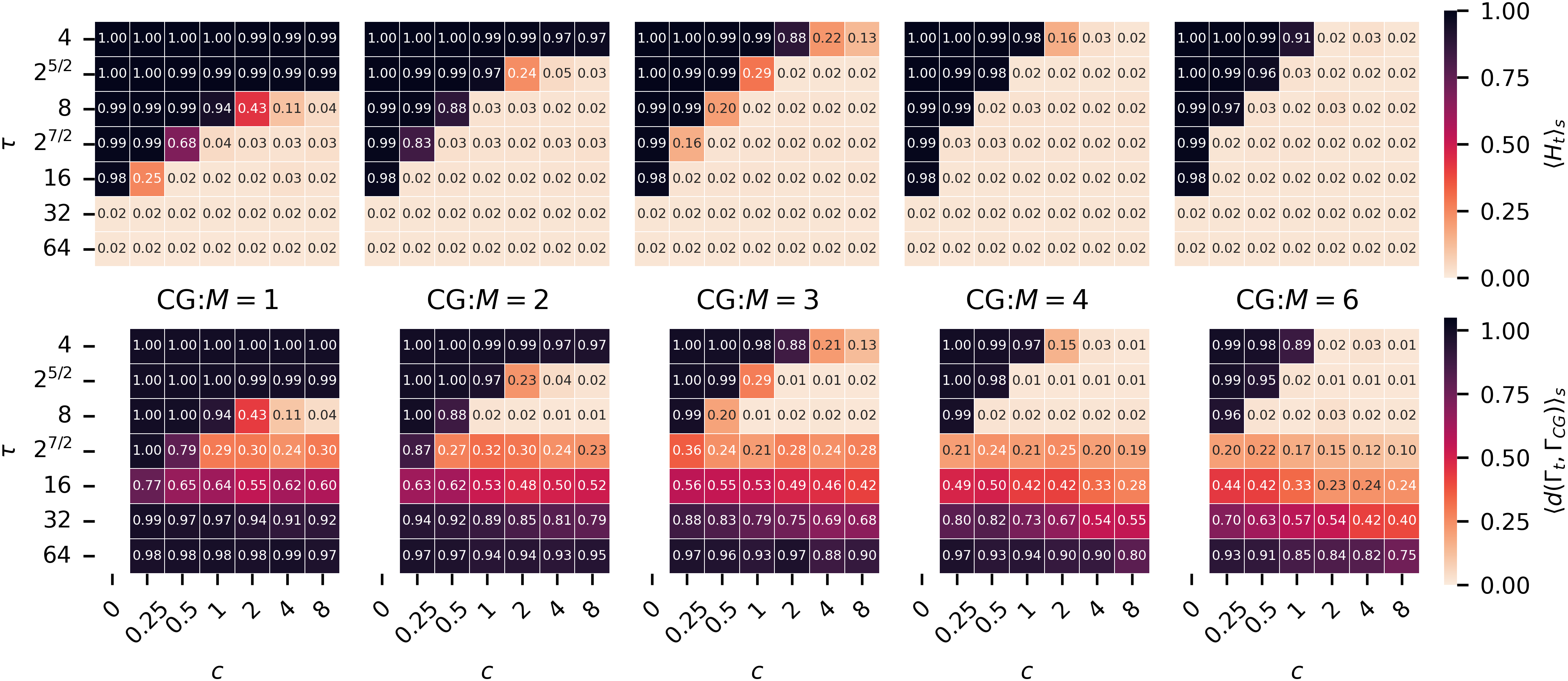}
    \caption{Heatmaps reporting stable states of the average entropy rate $\savg{H_\loops}$ (top) and the distance from the CG target path $\savg{ d\left(\Gamma_\loops,\Gamma_{CG}\right)}$ (bottom) as a function of decay time $\tau$ and meta-reinforcement contribution $c$ for different CG scales, $p_{sup}=0.999$.
    }
    \label{fig:CG_steady_rescaled_P_0.999_T_100}
\end{figure*}
Steady-state results for entropy and path distance are reported in Fig.~\ref{fig:CG_steady_rescaled_P_0.999_T_100}.
Together, they show that increasing  $M$ broadens the region of the parameter space associated to effective memorisation, without changing the qualitative behaviour of the system.
In particular, we highlight how the entropy rate $\savg{H_\loops}$ shows the same lowest values regardless of $M$. Similarly $\savg{ d\left(\Gamma_{\loops},\Gamma_{CG}\right)}$ reaches the same lowest value, $0.01$, when $M \geq 2$.
For $M=1$, instead, the system displays memorisation 
only for $\tau=2^{\nicefrac{7}{2}},8$ and the larger values of $c$.

\begin{figure}[t]
    \centering\includegraphics[width=0.9\columnwidth]{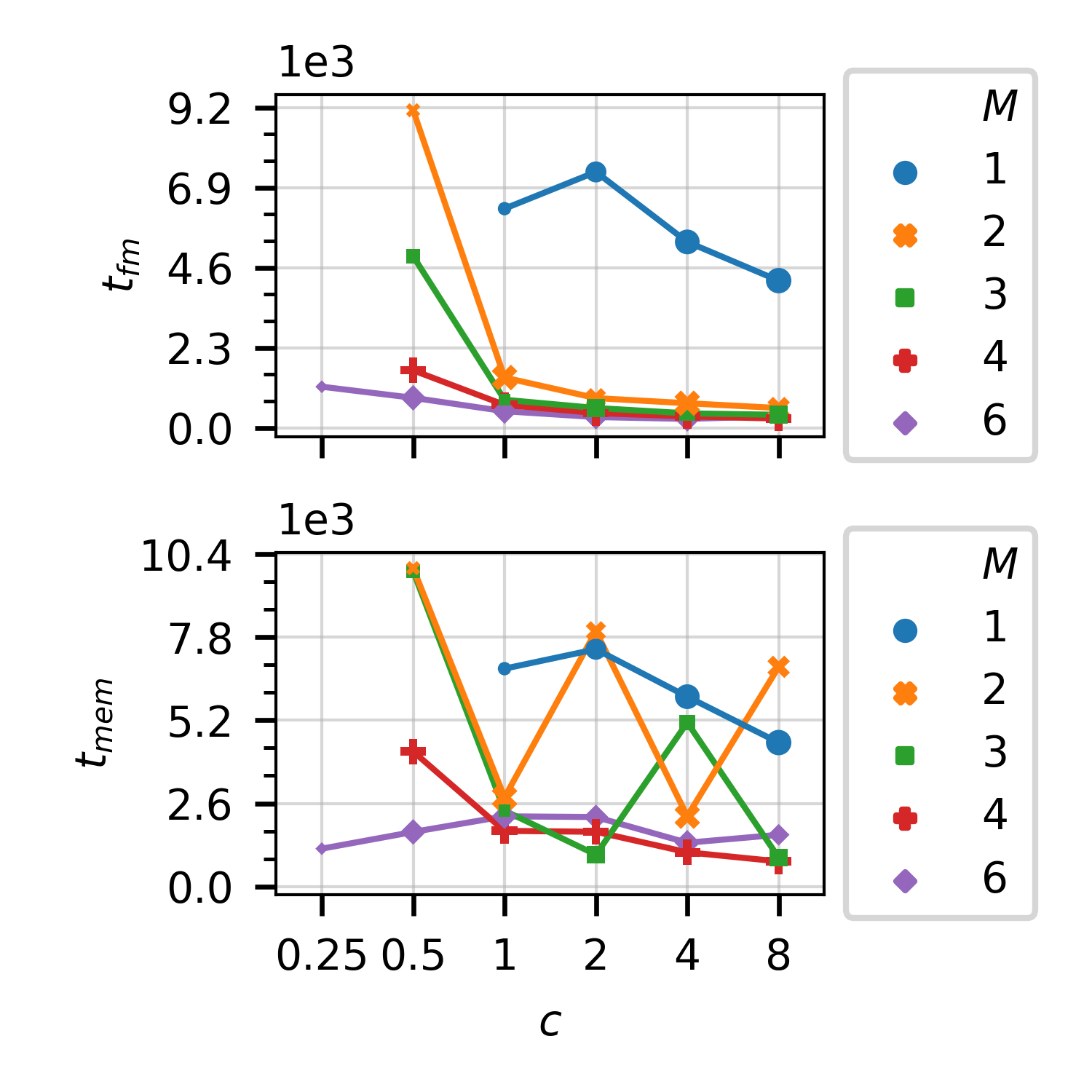}
    \caption{Average time $\loops_{fm}$ to first reach a memorising state (top) and memorisation time $\loops_{mem}$ as functions of the meta-reinforcement contribution $c$ (in $\log_{2}$-scale).
    Different curves correspond to different meta-reinforcement scale $M$ (colours and bullet shapes) for $\tau = 8$, $\prob_{sup}=0.999$.
    Bullet sizes reflect the proportion of memorising runs.}
    \label{fig:CG_arival_tau8}
\end{figure}
Focusing on $\tau=8$, Fig.~\ref{fig:CG_arival_tau8} shows the results for $\loops_{fm}$ and $\loops_{mem}$ as function of the meta-reinforcement contribution $c$ ($x$-axis). The results indicate that $\loops_{fm}$ decreases with increasing $M$, while the number of memorizing runs increases, in line with the results reported in Fig.~\ref{fig:CG_steady_rescaled_P_0.999_T_100}.
Furthermore, for $M \geq 2$ $\loops_{fm}$ decreases monotonically with $c$, also in agreement with Fig.~\ref{fig:CG_steady_rescaled_P_0.999_T_100}.
We attribute the non-monotonic behaviour of the $M=1$ line to the poor statistics of the $c=1$ case, for which the number of successful runs is smaller.
For $\loops_{mem}$ on the other hand, we see the non-monotonic scaling in $c$ we saw in Fig.~\ref{fig:memorisation_times_k0_100_T_100}
, which again seems to indicate the appearance of meta-stable states of $H_t$.

The above results can be understood in terms of information required to describe a CG path. Increasing $M$ increases the number of paths on the network that are equivalent to the CG path to be memorised, thus making it easier to recover it. 
Still, the fact that the lowest value of $\savg{H_t}$ is the same irrespective of $M$ shows that the behaviour of the system is not trivial. In fact this suggests that, even though all exact paths inside the CG path are theoretically equally valid, only one is memorised.

\section{Asymptotic behaviour of meta-reinforcement dynamics}
\label{sec:discussion}
\begin{figure}[ht]
    \centering
    \includegraphics[width=0.85\columnwidth]{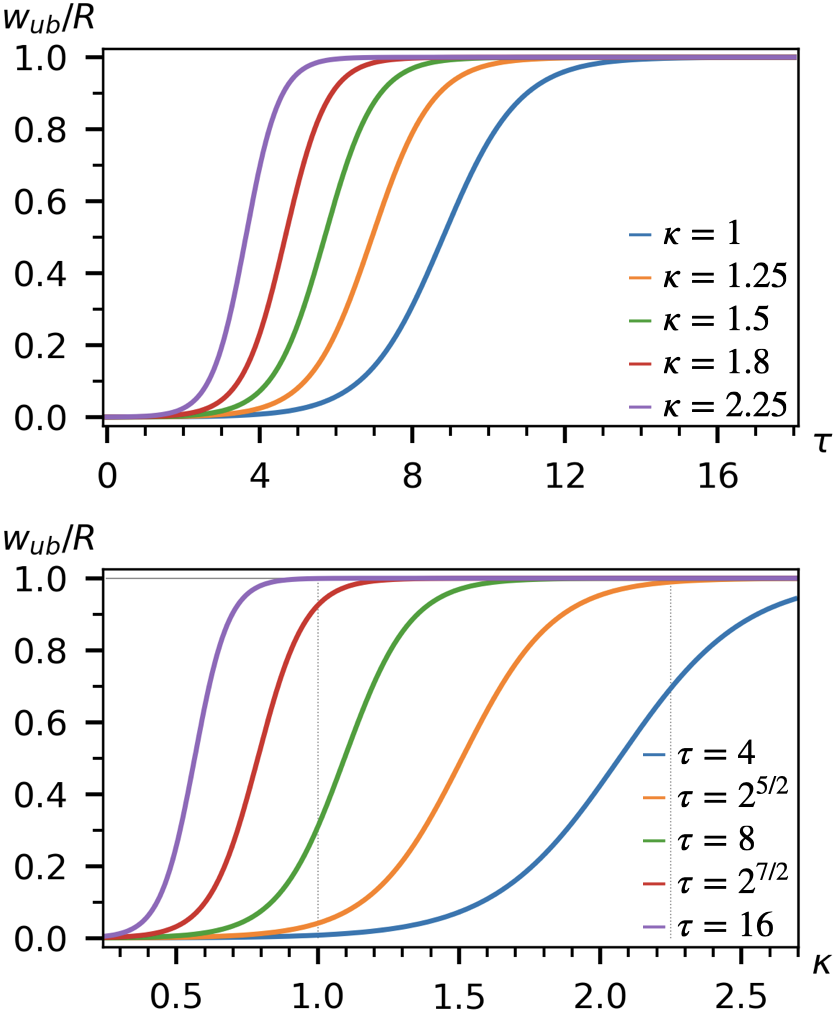}
    \caption{Normalised logistic functions $\nicefrac{w_{ub}}{R}$.
    Top: $\nicefrac{w_{ub}}{R}$ as a function of $\tau$ for different values of $\kappa^{\edge}$.
    Bottom: $\nicefrac{w_{ub}}{R}$ as a function of $\kappa^{\edge}$ for different values of $\tau$.
    Dashed vertical lines represent the initial steepness $\kappa=1$ and the maximal one achievable through meta-reinforcement $\kappa=2.25$.}
    \label{fig:weight_upperbounds}
\end{figure}
The interplay between meta-reinforcement and reinforcement can be understood more easily by considering the limit case in which a given edge in a group is always traversed, while all other edges in the same group are never chosen by the walker. This assumption is supported by the fact that all memorising systems show the same value of $\savg{H_\loops}$ irrespective of $M$.

As described in section~\ref{sec:methods}, our model relies on a ``hidden variable'', $s$, which influences the RW through the edge weight $w$.
Under the assumption that an edge is always selected, $s$ converges to a limit value $s_{\infty}$
\begin{equation}
\label{eq:traversal_function_upper_bound}
    s_{\infty} = \frac{1}{1-e^{-\frac{1}{\tau}}},
\end{equation}
as can be found by imposing $s_{\loops+1} = s_{\loops}$ in Eq.~\ref{eq:frecency}, or by realising that $s_{\loops}$ can be expressed as a geometric series in $e^{-\frac{1}{\tau}}$. 
Expanding Eq.~\ref{eq:traversal_function_upper_bound} around $\frac{1}{\tau}\sim 0$, one gets
\begin{equation}
\label{eq:s_ub_appr}
    s_{\infty} \simeq \tau + \frac{1}{2} + \frac{1}{12\tau} + o(\tau^{-2})
\end{equation}
so that for $\tau \gg 1$, $s_\infty \sim \tau +\frac{1}{2}$.
Substituting Eq.~\ref{eq:traversal_function_upper_bound} into Eq.~\ref{eq:sigmoid_logistic_function}, one gets an upper bound $w_{ub} < R$ for the maximum value reachable by the weights, $w_{ub} = \sigma\left(s_\infty\right)$. Using Eq.~\ref{eq:s_ub_appr}, this can be approximated as:
\begin{equation}
\label{eq:W_ub_appr}
    w_{ub} \simeq \frac{R}{1+e^{-\kappa\tau -\frac{\kappa}{2}+\xi}}. 
\end{equation}
$w_{ub}$ can thus be considered as a logistic function of either the learning rate $\kappa$ or $\tau$, whose upper bound is set by $R$. Due to the factor $\kappa\tau$, increasing $\tau$ leads to steeper functions of $\kappa$ and vice versa, see Fig.~\ref{fig:weight_upperbounds}.

It is important to note that the weight of any edge $\edge$ on the network approaches the maximum value $w_{ub}$ as soon as $\edge$ is traversed for $\tau$ consecutive steps, and decreases again when the edge is not selected. For $\prob_{sup}\sim 1$, however, consecutive traversals are probable and choosing a different path becomes more difficult, so that we  get some insight on the results reported before for $\prob_{sup} = 0.999$ by studying the behaviour of $w_{ub}\left(\kappa\right)$ at fixed $\tau$, reported in Fig.~\ref{fig:weight_upperbounds} (bottom panel). 

During our simulations, $\kappa$ is increased for meta-reinforced edges, going from a value $\kappa_0 = 1$ to up to $\kappa = 2.25$ depending on the coefficient $c$. 
Fixing $\kappa=1$ in Eq.~\ref{eq:W_ub_appr}, we get that the approximation to $w_{ub}\left(\kappa=1\right)/R$ is equal to $\frac{1}{2}$ for $\tau^*=\xi-\frac{1}{2}$. 
Substituting the value of $\xi$ used in this study, this gives $\tau^*=8.6$.
With this value of $\xi$ we further get $w_{ub}\left(\kappa=1\right)/R\geq 0.99$ for $\tau \geq 13$. Instead, for small values of $\tau$ $w_{ub}(\kappa=1)/R\sim 0$, and $w_{ub}\left(\kappa=2.25\right)<R$.
We can thus identify three regimes of interplay between reinforcement and meta-reinforcement in the limit case.
For $\tau\gtrsim 13$, reinforcement dominates, as increasing $\kappa$ leads to a negligible increase in $w_{ub}$.
For $\tau\lesssim 6$, meta-reinforcement can dominate, but only if $\kappa$ is increased rapidly, i.e. for $c>1$.
For $\tau\sim 8$, instead, increasing $\kappa$ considerably increases $w_{ub}$, leading to meta-reinforcement dominating over reinforcement also for relatively small values of $c$.

The limit regimes agree well with the behaviour observed in Sec.~\ref{sec:results} for the system with $p_{sup}=0.999$, both for memorisation and recall.
In the first case, the results in Fig.~\ref{fig:heatmap_steady_k0_100_T_100} show that although the system reaches low entropy rate values for $\tau \geq 16$ the distance from the meta-reinforced path remains high, indicating that reinforcement is dominating.
For $\tau=4$, the meta-reinforced path is practically never recovered, even when $c=8$.
Instead, $\tau = 2^{\nicefrac{5}{2}}, 8$ correspond to a sweet-spot, minimising both entropy rate and distance.
A similar behaviour can be observed in Fig.~\ref{fig:recall_scatters_k0_100_T_100}, with $\tau=8$ being the optimum and $\tau=2^{\nicefrac{5}{2}}$ competing with it only when $c=4,8$. 
The same figures also show that the network dynamics is richer than the asymptotic behaviour, which loses strength when $p_{sup}$ is lowered.
In that case, the interplay between reinforcement and meta-reinforcement should take into account also the average probability of choosing a reinforced edge, leading to a shift in the optimal value of $\tau$.
Intuitively, larger values of $\tau$ are needed to increase the probability of traversing the same edge $s_\infty$ times. 
This analysis is left to a future work.


\section{Conclusion}
\label{sec:conclusion}
In this manuscript we presented a simple model for a multi-level adaptive network whose connections follow a short-term Hebbian dynamics modulated by a long-term meta-plastic mechanism. The latter has the effect of changing the ``learning rate'' with which different connections are reinforced.
Our model describes phenomena in which a system of agents on a network adapt the topology of the latter to store information. This includes the case in which the agent is actually a walker exploring different locations.
For simplicity, we implemented the dynamics as a biased random walk on a cyclic feed-forward network.
While the model can be generalised to any strongly connected network topology, this choice allowed us to easily compute the entropy rate and distance between paths, facilitating the analysis of the dynamics. 

To the best of our knowledge, this is the first study investigating the dynamics of a system where plastic reinforcement is supported by a meta-plastic term changing the ``learning rate'' of selected connections.

By numerically exploring different parameters for the system, we have shown that there is a non-trivial interplay between plastic and meta-plastic reinforcements, depending on the decay time, $\tau$, of the first and the rate, $c$, with which the second acts.
We identified three regimes.
In the first, for large values of $\tau$, reinforcement dominates, and the long-term effect of meta-reinforcements has negligible influence.
For small values of $\tau$, the system starts showing memory only if the meta-reinforcement rate $c$ is large.
Interestingly, we also identified an intermediate regime in which even a small amount of meta-reinforcement can modulate memory formation.
Above all, we showed that in the latter two regimes meta-reinforcement allows both the storing of a path, and its retrieval when all weights have been reset to their initial value, i.e., when Hebbian memory has been erased.
Finally, we studied various CG scales for meta-reinforcement. Our results show that the qualitative behaviour of the system is robust at all investigated scales, with coarser ones facilitating the formation and recovery of memorised CG paths.

We rationalised our results considering the limit case in which a single edge in a group is always selected, maximally reinforcing a weight. We show that the maximum value of the weight grows as a sigmoid whose steepness is controlled by $\tau$ and the meta-reinforcement term $\kappa(c)$, allowing one to recover the three regimes by comparing the values of $\tau$ and $c$. 

The model presented in this work is inspired by brain dynamics and features but it is completely agnostic of the biological reality and suited to describe any two-scale memorisation process.
Therefore, we believe our model to be applicable to any biological and social system and dynamics that can be mapped onto an adaptive network with multiple distinct memory contributions, particularly opinion diffusion, epidemics modelling and ecological networks.

\section*{Acknowledgements}
We thank Dr. Beatrice Vignoli and Prof. Marco Canossa for helpful insights in the biological aspects of memory for the development of the model;
Dr. Alessandro Ingrosso and Dr. An\v{z}e Bo\v{z}i\v{c} for useful suggestions and discussions.
%
%
\section*{Funding}
This research was supported by the European Research Council (ERC) under the European Union’s Horizon 2020 research and innovation program [grant agreement No. 788793, BACKUP]. 
\section*{Author contributions}
LT, PB and LP conceived the work. LT, GZ, and PB devised a preliminary version of the model. GZ developed the model and the data analysis techniques with help from LT, PB and JM. GZ carried out simulations, cleaned and analysed data. All authors reviewed the results and contributed to their interpretation. GZ wrote the first draft of this manuscript. All authors contributed to and approved the final version of this manuscript.
%
\bibliographystyle{apsrev4-2}
\bibliography{phdbib}
\clearpage
\onecolumngrid
\appendix*
\renewcommand\thefigure{S\arabic{figure}}
\setcounter{figure}{0} 
\section*{Supplementary Material}
\label{sec:supplementary}
%
%
\begin{figure*}[!htb]
    \centering
    \includegraphics[width=0.9\textwidth]{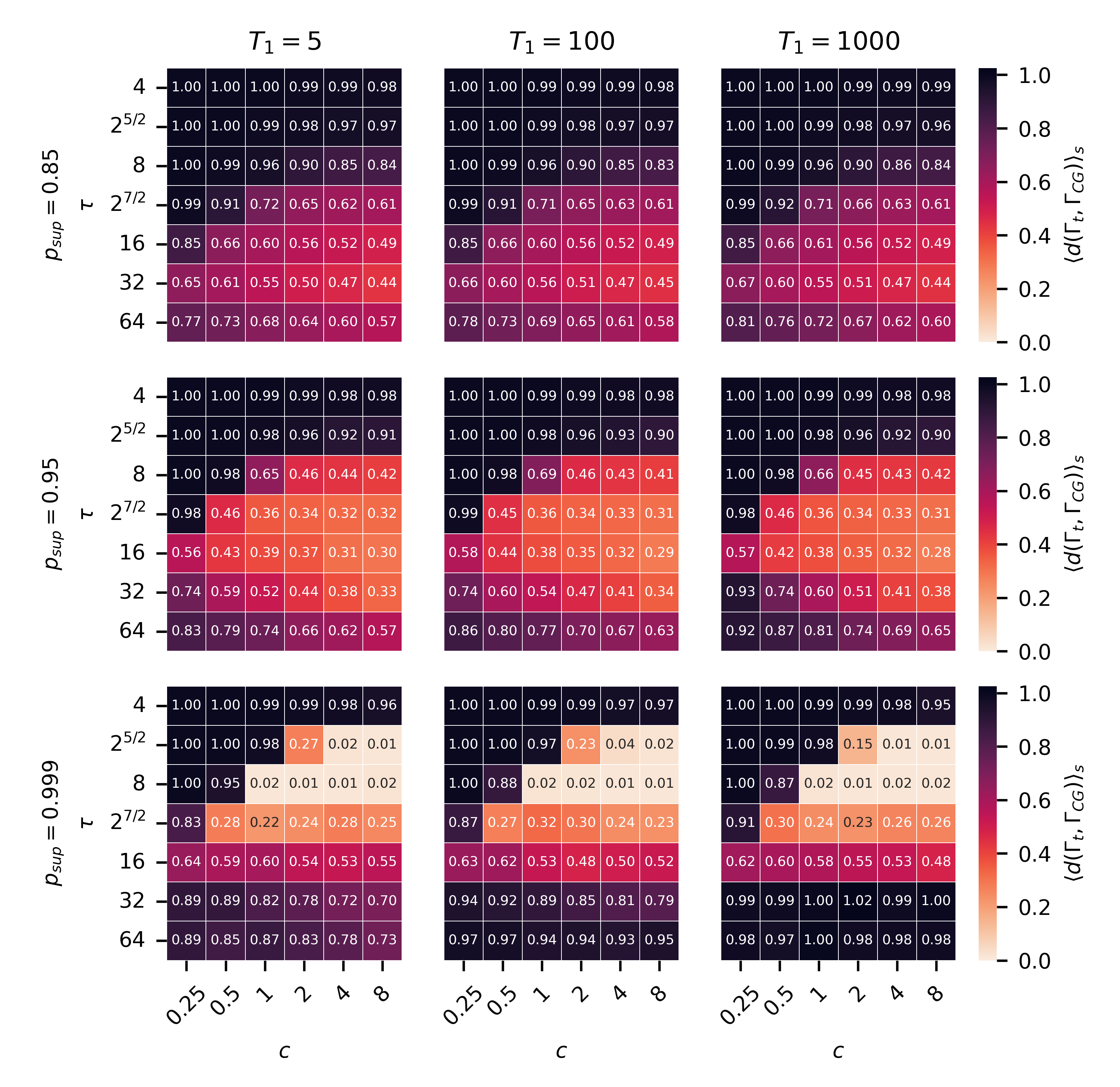}
    \caption{Distance heatmaps for all values of $T_{1}$.
    Heatmaps reporting stable states of the average distance from the coarse-grained target path $\savg{d \left(\Gamma_{\loops},\Gamma_{\mathrm{CG}}\right)}$ as a function of decay time $\tau$ and meta-reinforcement contribution $c$ for different values of $\prob_{\mathrm{sup}} \in \{ 0.85, 0.95, 0.999\}$.
    Condition $c=0$ is not reported, as in absence of meta-reinforcement the distance from target path $\Gamma_{\mathrm{CG}}$ is uninformative.
    Corresponding values for the entropy rate $\savg{H_{\loops}}$ are not reported because they are not affected by parameter $T_{1}$ provided that $T_{2}+T_{3} \gg T_{1}$, which is true in our case.}
    \label{fig:heatmaps_steady_k0_100_allT}
\end{figure*}
\begin{figure}[t]
    \centering
        \subfloat{%
          \includegraphics[width=0.475\textwidth]{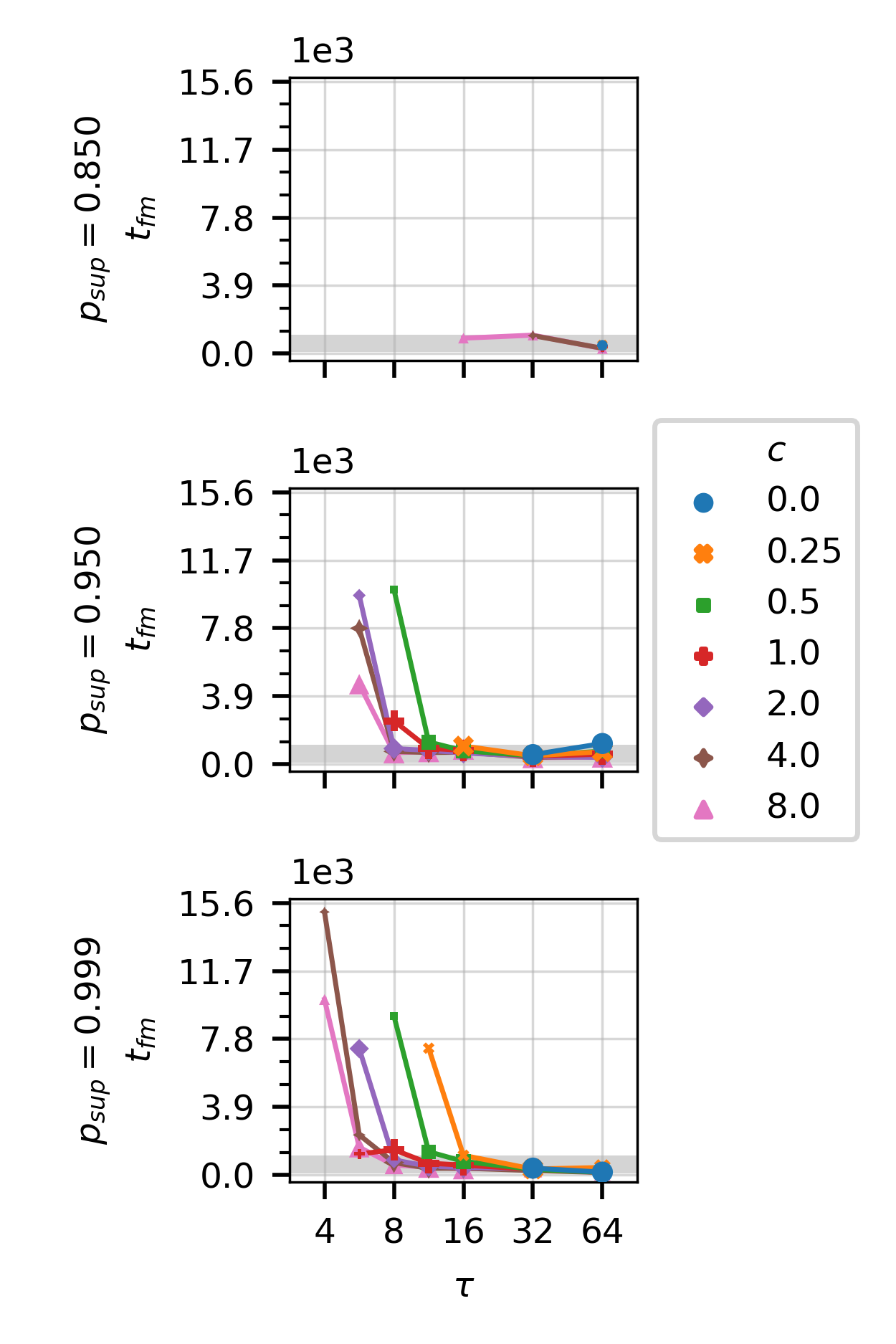}%
        }
        \hfill
        \subfloat{%
          \includegraphics[width=0.475\textwidth]{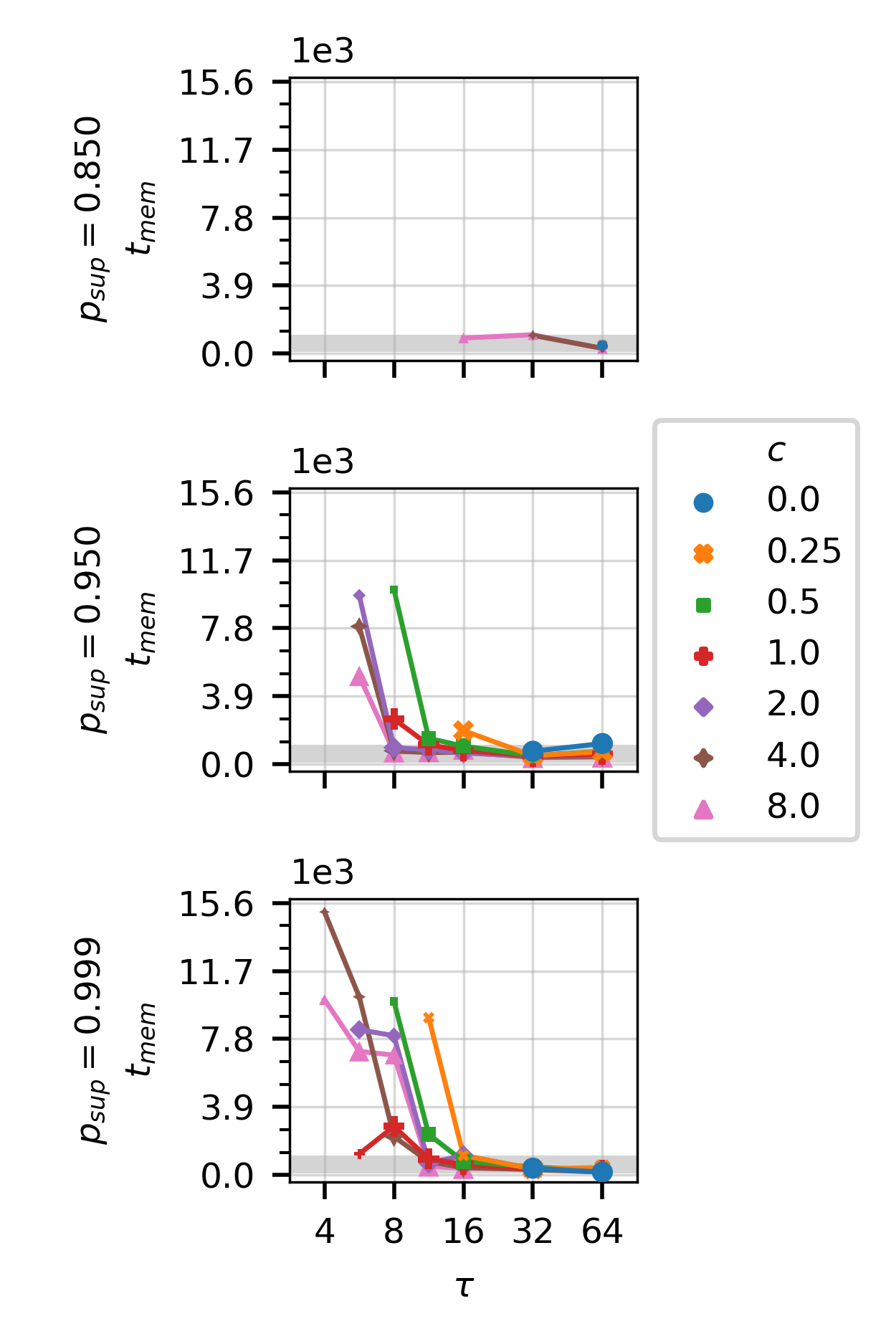}%
        }
    \caption{Average time $\loops_{\mathrm{fm}}$ to drop below the memorisation threshold (left) and memorisation time $\loops_{\mathrm{mem}}$ (right) as functions of the decay times $\tau$ (in $\log_{2}$-scale) for different values of $\prob_{\mathrm{sup}} \in \{ 0.85, 0.95, 0.999\}$ and $T_{1}=100$.
    Different curves correspond to different values of the meta-reinforcement contribution $c$ (colours and bullet shapes).
    Values are averaged over runs that did reach a state below $H^{*}$:
    bullet sizes reflect the amount of these runs, out of 100.
    The grey shaded area indicates the meta-reinforcement window from $\loops=T_{1}$ to $\loops=T_{1}+T_{2}$.}
    \label{fig:memorisation_times_k0_100_T_100_allP}
\end{figure}
%
%
\begin{figure*}[t]
    \centering
    \includegraphics[width=0.9\textwidth]{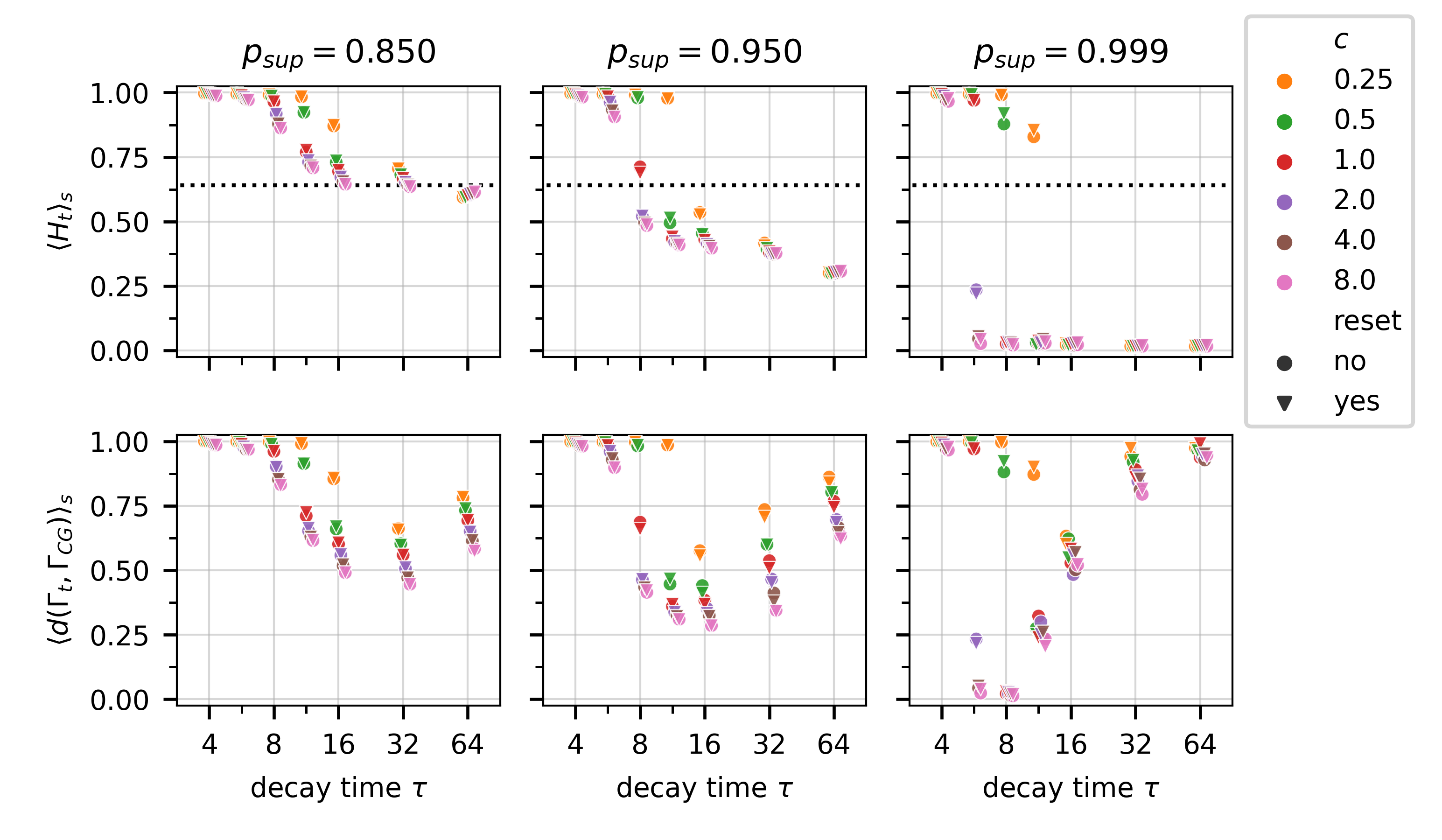}
    \caption{Recall analysis through entropy rate $\savg{H_{\loops}}$ (above) and distance from the target CG path $\savg{d\left( \Gamma_{\loops},\Gamma_{\mathrm{CG}} \right)}$ (below) as functions of the decay times $\tau$ ($\log_{2}$ scale), the meta-reinforcement coefficient $c$ (colours) for different values of $\prob_{\mathrm{sup}} \in \{ 0.85, 0.95, 0.999\}$ and $T_{1}=100$.
    Symbols are slightly displaced along the $x$-axis to ease visualisation and comparison. Bullet shape reflects when the steady state is computed:
    before (circular bullets) or after (triangular bullets) the reset.
    The dotted line in the top panels represents the threshold $H^{*}$ associated to memorisation.}
    \label{fig:recall_scatters_k0_100_T_100_allc}
\end{figure*}
%
%
\begin{figure*}[ht!]
    \centering
    \includegraphics[width=\textwidth]{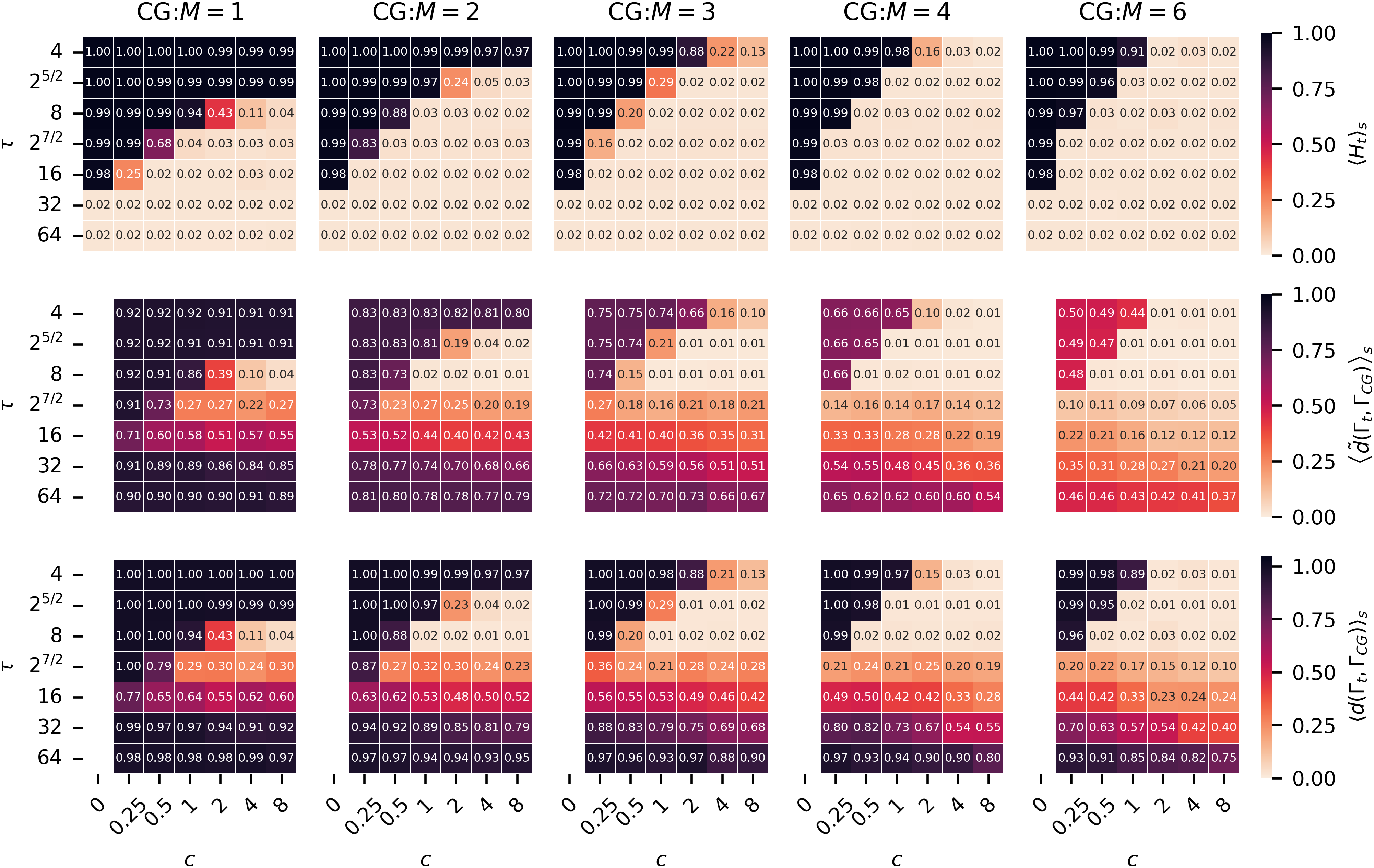}
    \caption{Heatmaps reporting stable states of entropy rate $\savg{H_{\loops}}$ (top) and distance $\savg{d \left( \Gamma_{\loops},\Gamma_{\mathrm{CG}}\right)}$ (bottom) from the CG target path $\Gamma_{\mathrm{CG}}$, together with a non-rescaled (i.e. without the $\frac{N}{N-M}$ factor) distance $\savg{\tilde{d} \left( \Gamma_{\loops}, \Gamma_{\mathrm{CG}}\right)}$ (middle), as functions of decay time $\tau$ and meta-reinforcement contribution $c$ for different CG scales of the meta-reinforcement dynamics, $\prob_{\mathrm{sup}}=0.999$ and $T_{1}=100$.
    The columns corresponding to $c=0$ in middle and bottom heatmaps are not reported as in absence of meta-reinforcement the distance from target path $\Gamma_{\mathrm{CG}}$ is uninformative.}
    \label{fig:CG_steady_gloabl_P_0.999_T_100}
\end{figure*}
\begin{figure*}[ht!]
    \centering
    \includegraphics[width=\textwidth]{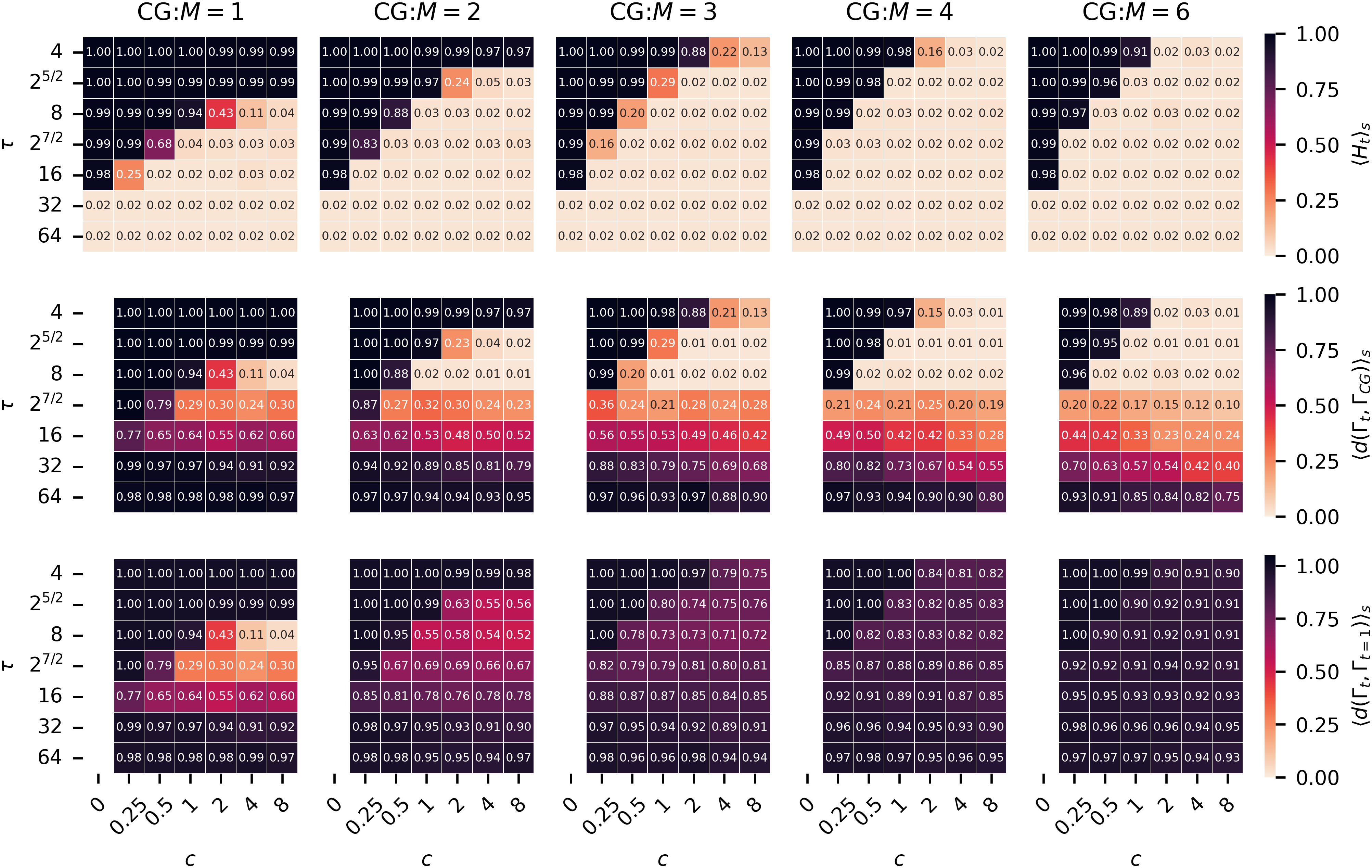}
    \caption{Heatmaps reporting stable states of entropy rate $\savg{H_{\loops}}$ (top) and distances from the CG target path $\savg{d\left( \Gamma_{\loops}, \Gamma_{\mathrm{CG}}\right)}$ (middle) and from the ``exact network target path'' $\savg{d\left( \Gamma_{\loops}, \Gamma_{\loops=1} \right)}$ as a function of decay time $\tau$ and meta-reinforcement contribution $c$ for different CG scales of the meta-reinforcement dynamics, $\prob_{\mathrm{sup}}=0.999$ and $T_{1}=100$.
    The columns corresponding to $c=0$ in middle and bottom heatmaps are not reported, as in absence of meta-reinforcement the distance from target path $\Gamma_{\mathrm{CG}}$ is uninformative.}
    \label{fig:CG_steady_withnodes_P_0.999_T_100}
\end{figure*}
\begin{figure*}[ht!]
    \centering
    \includegraphics[width=\textwidth]{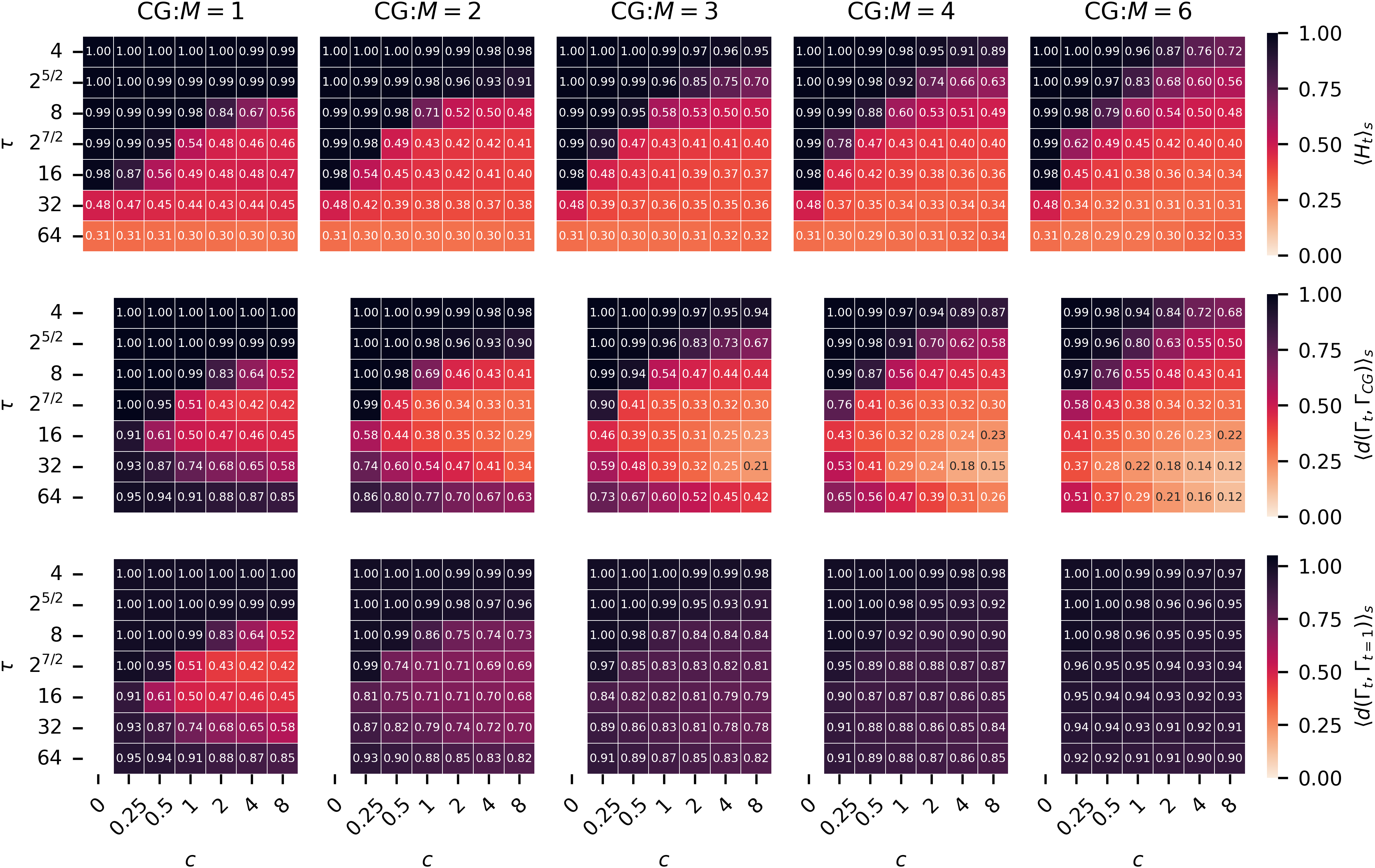}
    \caption{Heatmaps reporting stable states of entropy rate $\savg{H_{\loops}}$ (top) and distances from the CG target path $\savg{d\left( \Gamma_{\loops}, \Gamma_{\mathrm{CG}}\right)}$ (middle) and from the ``exact network target path'' $\savg{d\left( \Gamma_{\loops}, \Gamma_{\loops=1} \right)}$ as functions of decay time $\tau$ and meta-reinforcement contribution $c$ for different CG scales of the meta-reinforcement dynamics, $\prob_{\mathrm{sup}}=0.950$ and $T_{1}=100$.
    The columns corresponding to $c=0$ in middle and bottom heatmaps are not reported, as in absence of meta-reinforcement the distance from target path $\Gamma_{\mathrm{CG}}$ is uninformative.}
    \label{fig:CG_steady_withnodes_P_0.950_T_100}
\end{figure*}
\begin{figure*}[ht!]
    \centering
    \includegraphics[width=\textwidth]{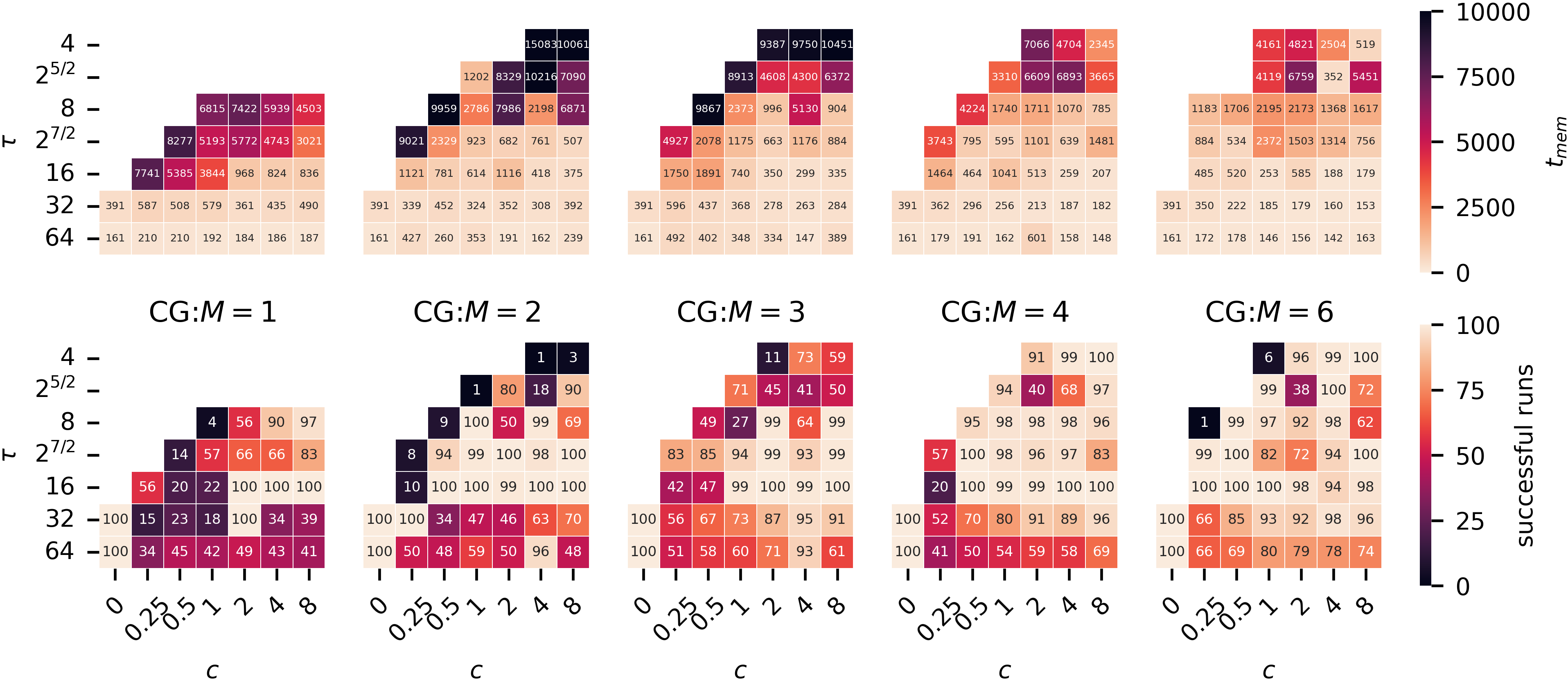}
    \caption{Heatmaps reporting the average memorisation time $\loops_{mem}$ (top) and the number of runs showing successful memorisation (bottom) as a function of decay time $\tau$ and meta-reinforcement contribution $c$ for different CG scales of the meta-reinforcement dynamics, $\prob_{\mathrm{sup}}=0.999$ and $T_{1}=100$.}
    \label{fig:CG_transitions_P_0.999_T_100}
\end{figure*}

\end{document}